\documentclass[review]{elsarticle}

\usepackage{epsfig}

\usepackage{amssymb}
\usepackage{amsthm}
\usepackage{amsmath}
\usepackage{xcolor}
\usepackage{array}
\usepackage{comment}
\usepackage{multirow}
\newdimen\figrasterwd
\figrasterwd\textwidth

\theoremstyle{definition}
\newtheorem{def1}{Definition}
\newtheorem{def2}[def1]{Definition}
\newtheorem{def3}[def1]{Definition}
\newtheorem{def4}[def1]{Definition}
\newtheorem{def5}[def1]{Definition}
\newtheorem{def6}[def1]{Definition}
\newtheorem{exp2}{Example}
\newtheorem{exp1}[exp2]{Example}
\newtheorem{rmk1}{Remark}
\newtheorem{rmk2}[rmk1]{Remark}
\newtheorem{rmk3}[rmk1]{Remark}
\newtheorem{rmk4}[rmk1]{Remark}
\newtheorem{rmk5}[rmk1]{Remark}
\newtheorem{rmk6}[rmk1]{Remark}
\newtheorem{rmk7}[rmk1]{Remark}
\newtheorem{rmk8}[rmk1]{Remark}
\newtheorem{rmk9}[rmk1]{Remark}
\newtheorem{rmk10}[rmk1]{Remark}
\newtheorem{rmk11}[rmk1]{Remark}

\theoremstyle{plain}
\newtheorem{thm1}{Theorem}
\newtheorem{thm2}[thm1]{Theorem}
\newtheorem{thm3}[thm1]{Theorem}
\newtheorem{prop1}{Proposition}

\newtheorem{prop3}{Lemma}

\newtheorem{prop5}[prop3]{Lemma}
\newtheorem{lem4}[prop3]{Lemma}
\newtheorem{lem5}[prop3]{Lemma}

\journal{}

\usepackage{cite}
\usepackage{amsmath,amssymb,amsfonts}
\usepackage{graphicx}
\usepackage{algorithm,algorithmic}
\usepackage{array}
\usepackage[caption=false,font=normalsize,labelfont=sf,textfont=sf]{subfig}
\usepackage{textcomp}
\usepackage{stfloats}
\usepackage{url}
\usepackage{verbatim}
\usepackage{textcomp}
\usepackage{appendix}

\begin{document}


\begin{frontmatter}

\title{Distributed multi-UAV shield formation based on virtual surface constraints}

\author[inst1]{María Guinaldo\corref{mycorrespondingauthor}}
\cortext[mycorrespondingauthor]{Corresponding author}
\ead{mguinaldo@dia.uned.es}

\affiliation[inst1]{organization={Department of Computer Science and Automatic Control of Universidad Nacional de Educación a Distancia (UNED)},
            addressline={Juan del Rosal 16}, 
            city={Madrid},
            postcode={28040}, 
            country={Spain}}

\author[inst1]{José Sánchez-Moreno}

\author[inst2]{Salvador Zaragoza}

\author[inst1]{Francisco José Mañas-Álvarez}

\affiliation[inst2]{organization={Centro Universitario de la Defensa (CUD)},
            addressline={Coronel López Peña s/n}, 
            city={Santiago de la Ribera, Murcia},
            postcode={30729}, 
            country={Spain}}

\begin{abstract}
This paper proposes a method for the deployment of a multi-agent system of unmanned aerial vehicles (UAVs) as a shield with potential applications in the protection of infrastructures. 
The shield shape is modeled as a quadric surface in the 3D space. To design the desired formation (target distances between agents and interconnections), an algorithm is proposed where the input parameters are just the parametrization of the quadric and the number of agents of the system. This algorithm guarantees that the agents are \textit{almost uniformly} distributed over the \textit{virtual} surface and that the topology is a Delaunay triangulation. Moreover, a new method is proposed to check if the resulting triangulation meets that condition and is executed locally. Because this topology ensures that the formation is rigid, a distributed control law based on the gradient of a potential function is proposed to acquire the desired shield shape and proofs of stability are provided. Finally, simulation and experimental results illustrate the effectiveness of the proposed approach.
\end{abstract}




\end{frontmatter}


\section{Introduction}
\label{sec:introduction}
The use of autonomous robot systems that work cooperatively for different tasks related to robotics has been growing in the last few years. 
The deployment of a formation is used, for instance, in sampling, monitoring, or surveillance tasks \citep{Leonard2007,Fidan2007,Aranda2015}.
In this context, each entity of the system is also called \textit{agent} and the system is referred to as \textit{multi-agent}. In all the aforementioned tasks, maintaining a formation of the robots plays a crucial role, and the design of distributed control laws that guarantee the achievement and maintenance of such objective is an active line of research \citep{Fredslund2002,Lawton2003}. 

Different proposals exist depending on the agents' measurement capabilities and the assumptions that are taken \citep{Oh2015}. On the one hand, regulating the relative position of pairs of agents \citep{Olfati2004} allows simpler control algorithms and stability analysis but requires the agents to have a common global coordinate frame or local coordinate frames with the same orientation. On the other hand, if the formation is defined in terms of target distances between pair of agents \citep{Krick2009,Cao2011}, the control law can be computed with respect to the agent's local frame, which does not need to have a common orientation, although ambiguities in the positioning of the agents \citep{Kwon2022} or non-robust behaviors \citep{Mou2015} can occur. In this regard, graph rigidity has allowed the design of distributed control laws for formation control that reduce these ambiguities \citep{Anderson2008,Marina2014}. These are usually based on the gradients of the potential functions closely related to the graphs describing the distance constraints between the neighboring agents. 

Related to the concept of rigidity, a Delaunay triangulation belongs to the class of proximity graphs \citep{Mathieson2019}, and it is the dual of the Voronoi Diagram \citep{Hjelle2006}. The graph of a Delaunay triangulation is rigid (but not minimally rigid in general), and then, the associated formation is stable, at least locally. In this regard, the existence of multiple equilibria of the potential function adds considerable complexity to the convergence analysis of formation control algorithms \citep{Sun2015}, and only strong results have been obtained for relatively simple settings in 2D \citep{Dorfler2010,Anderson2010,Fathian2019}, and global stabilization of rigid formation in arbitrary dimensional spaces still remains an open problem. Moreover, as reported in \citet{Krick2009}, when the formation is not minimally rigid, the extra edges might cause the system to have additional equilibrium points. 
Recently, some strategies have been proposed by introducing extra variables such as angles \citep{Liu2020} or areas \citep{Anderson2017} in the constraints to reduce the number of possible non-desired equilibria, allowing the expansion of the region of attraction of the desired equilibrium set. However, more sophisticated equipment might be required to measure new variables, and, in case of inconsistent measures \citep{Marina2014}, the possibility of undesired behavior increases. Moreover, tight constraints are imposed on the graph that describes the triangulation, for instance, the graph is restricted to be a leader-first-follower (LFF) minimally persistent directed graph \citep{Summers2011}, which restricts the out-degree to 2.

A 2D scenario is not applicable when agents are aerial robots or drones, which move in the 3D space, and in this case, the existing results on rigid formations are scarce. In \citet{Brandao2016}, a multi-layer control scheme for positioning and trajectory tracking missions in UAVs is presented. A Delaunay triangulation is used to decompose the group of UAVs into triangles, which are guided individually by a centralized and multi-layer controller. In \citet{Park2014} a tetrahedral shape formation of four agents is studied. In \citet{Ramazani2016} a 3D setting is proposed in which a subset of agents are constrained to move in a plane and form with the rest a triangulation that is minimally rigid. For a general state space, a control law is proposed in \citet{Park2017} that guarantees almost global convergence but requires the graph to be complete. The strategy of including additional constraints to reduce ambiguities \citep{Liu2020} has been extended to characterize a tetrahedron formation in 3D \citep{Liu2021}, and therefore has similar limitations to the 2D version regarding the graph, but with the out-degree constrained to 3. Finally, in \citep{Han2017}, a barycentric coordinate-based approach is proposed following a leader-follower approach allowing almost global convergence. However, a communication graph is introduced and an auxiliary state information is exchanged. Otherwise, a global optimization problem needs to be solved to compute feedback parameters \citep{Han2016}.

In this paper, we propose a strategy for the deployment of a formation of a group of UAVs modeled as single integrators around an area of interest. A potential application is the protection of infrastructures so that the multi-agent system would form a \textit{shield} to, for instance, the monitoring of external threats. For the control and maintenance of the formation, a distributed control law is proposed based on the gradient of a potential function that guarantees stability and the acquisition of the desired shield shape. In particular, the topology of the system modeled by a graph is a Delaunay triangulation and the shape of the shield is a quadric surface in the 3D space. Additionally, a simple procedure to design the target formation is presented: it only requires the quadric surface parameters and the total number of agents of the system, and as a result, an \textit{almost uniform} distribution of the agents over the surface and the desired topology are generated. 
Finally, and due to the fact that the shield is deployed in the 3D space, an extension of the local characterization of 2D Delaunay triangulations reported in \citep{Schwab2021} is proposed and applied with success to the quadratic surface to ensure that the resulting triangulation fulfills the required properties. We further validate our approach over an experimental platform of micro-aerial vehicles whose description can be found at \citep{Manas2023}. 

With respect to related work, the proposed strategy offers an integrated framework to both design the target formation and the control law to achieve it. On the one hand, the proposed algorithm to design the target formation uses a simple parametrization of the surface to compute the desired inter-distances between nodes so that an almost uniform distribution is achieved. The fact that no optimization problem is solved drastically reduces the computational cost, compared to traditional approaches in the plane in the context of ad-hoc networks \citep{Cortes2004}. Additionally, a new distributed method is proposed to check that the triangulation is Delaunay's in 3D surfaces since available results are restricted to the plane \citep{Schwab2021}. On the other hand, the existing literature on formation control strategies assumes that the parameters of the formation are given. Moreover, although recent works have addressed the shape control in 3D spaces \citep{Park2017}-\citep{Han2017}, to the best of the authors' knowledge, the proposed approach based on virtual surfaces embedded in the 3D space, has not been addressed. This constraint makes that the concept of infinitesimal rigidity \citep{Asimow1979} (which is the basis for many existing results) cannot be applied as such, and hence, new rigidity properties are derived to study stability, which is another contribution of the paper. Additionally, the proposed strategy is more flexible in the sense that it does not require a complete graph such as in \citep{Park2017} or out-degree constraints \citep{Liu2020,Liu2021}, which would not allow the deployment of a shield with a generic number of nodes $N$ and with a given shape. Also, communication is not required as in the barycentric approach \citep{Han2017}, and formation can achieved based on local measurements. Finally, although the number of indoor platforms with multi-agent aerial robots has been increasing in the last few years \citep{Chung2018,Fathian2019b}, still the validation of distance formation control strategies is mostly performed in simulation, and hence, the implementation of the approach over a team of 12 UAVs constitutes a challenge that has been addressed.

The rest of the paper is organized as follows: Section \ref{sec:preliminaries} introduces some preliminary concepts that will be used through the paper. Section \ref{sec:problemDescription} describes the problem to be solved in this paper. A simple procedure to define the target configuration is described in Section \ref{sec:ShieldBuilding}. The proposed control law and the stability analysis is provided in Section \ref{sec:controllaw}. Section \ref{sec:simulations} illustrates with simulations the results of the paper, and experimental results over a real testbed are also provided. Finally, Section \ref{sec:conclusions} provides the conclusions and future work.

\section{Preliminaries} \label{sec:preliminaries}

\subsection{Differential Geometry}
\begin{def1} A regular surface in Euclidean space $\mathbb{R}^3$ is a subset $S$ of $\mathbb{R}^3$ such that every point of $S$ has an open neighborhood $U \in \mathbb{R}^3$ for which there is a smooth function $F : U \to \mathbb{R}^2$ with:
\begin{itemize}
    \item $S \cap U = \{(x, y, z) \in U : \ F(x, y, z) = 0\}$.
    \item at each point of $S \cap U$, at least one partial derivative of $F$ is nonzero.
\end{itemize}
\end{def1}

We denote the Jacobian of a function $f: \mathbb{R}^n\to \mathbb{R}^m$ evaluated at a point $p$ as $J_f(p)$. In the special case when $f : \mathbb{R}^n\to \mathbb{R}$, the Jacobian of $f$ is the gradient of $f$ and we denote it by $\nabla f(p)$. Occasionally for convenience during
calculations of the Jacobian, the notation $\tfrac{\partial}{\partial p}$ will be
used to represent $J_f(p)= \tfrac{\partial}{\partial p}f(p)$.

\subsection{Graph theory} \label{sec:graphTheory}
Consider a set $\mathcal{N}$ of $N$ agents. The topology of the multi-agent system can be modeled as a static undirected graph $\mathcal{G}$. This section reviews some facts from algebraic graph theory \citep{Godsil2001}.
The graph $\mathcal{G}$ is described by the set of agent-nodes $\mathcal{V}$ and the set of edges $\mathcal{E}$.

For each agent $i$, $\mathcal{N}_i$ represents the neighborhood of $i$, i.e., $\mathcal{N}_i=\{j\in \mathcal{V}: \ (i,j)\in \mathcal{E}\}$. Note that $|\mathcal{N}_i|=\text{deg } v_i$, where $|\cdot|$ represents the cardinality of the set $\mathcal{N}_i$ and $\text{deg}$ is the degree of the vertex $v_i$ associated to the node $i$.

Assume that the edges have been labeled as $e_k$ and arbitrarily oriented, and its cardinality is labeled as $N_e$. Then the incidence matrix $H(\mathcal{G})=[h_{ik}]\in\mathbb{R}^{N\times N_e}$ is defined as $h_{ik}=-1$ if $v_i$ is the tail of the edge $e_k$, $h_{ik}=1$ if $v_i$ is the head of $e_k$, and $h_{ik}=0$ otherwise. The Laplacian matrix $L(\mathcal{G}) \in \mathbb{R}^{N\times N}$ of a network of agents is defined as $L(\mathcal{G})=H(\mathcal{G})H^\top(\mathcal{G})$. The Laplacian matrix $L(\mathcal{G})$ is positive semidefinite, and if $\mathcal{G}$ is connected and undirected, then $0=\lambda_1(\mathcal{G})<\lambda_2(\mathcal{G})\leq \dots \leq \lambda_N(\mathcal{G})$, where $\{\lambda_j(\mathcal{G})\}$ are the eigenvalues of $L(\mathcal{G})$. The adjacency matrix of $\mathcal{G}$ is $A(\mathcal{G})=[a_{ij}]$, where $a_{ij}=1$ if there is an edge between two vertices $v_i$ and $v_j$, and 0 otherwise. Matrices $H(\mathcal{G})$, $L(\mathcal{G})$ and $A(\mathcal{G})$ can be simply denoted by $H$, $L$ and $A$, respectively, when it is clear from the context.

\subsection{Graph rigidity} \label{sec:graphRigidity}
 
A framework is a realization of a graph at given points in Euclidean
space. We consider an undirected graph $\mathcal{G}=(\mathcal{V},\mathcal{E})$ with $N$ vertices embedded in $\mathbb{R}^m$, with $m=2$ or $m=3$ by assigning to each
vertex $i$ a location $p_i \in \mathbb{R}^m$. Define the composite vector $p = ( p_1, \ ..., \ p_n) \in \mathbb{R}^{mn}$. A framework is a pair $(\mathcal{G}, p)$. 

For every framework  $(\mathcal{G}, p)$, we define the rigidity function $f_{\mathcal{G}}(p): \mathbb{R}^{2N}\to \mathbb{R}^{N_e}$ given by
$$f_{\mathcal{G}}(p)=(\dots,\|z_k\|^2,\dots),$$
where $\|z_k\|^2=\|p_i-p_j\|^2$, corresponds to the edge $k$ in $\mathcal{E}$ that connects two vertices $i$ and $j$. Note that this function is not unique and depends on the ordering given to the edges.

The formal definition of rigidity and global rigidity can be found in \citet{Asimow1979}. But roughly speaking, a framework $(\mathcal{G}, p)$ is rigid if it is not possible to smoothly move some vertices of the framework without moving the rest while maintaining the edge lengths specified by $f_{\mathcal{G}}(p)$.

Let us take the following approximation of $f_{\mathcal{G}}(p)$:
$$f_{\mathcal{G}}(p+\delta p)=f_{\mathcal{G}}(p)+R(p)\delta p + O(\delta p^2),$$
where $R(p)=J_{f_\mathcal{G}}(p)$ denotes the Jacobian matrix of $f_{\mathcal{G}}(p)$, and $\delta p$ is an infinitesimal displacement of $p$. The matrix $R(p)$ is called the \textit{rigidity matrix} of the framework $(\mathcal{G}, p)$. Analyzing the properties of $R(p)$ allows to infer further properties of the framework. Next we present some existing results:

\begin{def2} \citep{Asimow1979}. A framekwork  $(\mathcal{G}, p)$ is infinitesimally rigid if rank$(R(p))=2N-3$ in $\mathbb{R}^2$ or rank$(R(p))=3N-6$ in $\mathbb{R}^3$. 
\end{def2}

Therefore, the kernel of $R(p)$ has dimension 3 and 6 in $\mathbb{R}^2$ and $\mathbb{R}^3$, respectively, which corresponds to the rigid body motions that makes that $R(p)\delta p =0$ with $\delta p\neq 0$. In $\mathbb{R}^2$, this corresponds to translation along $x$, translation along $y$ and the rotation about $z$. Similary, in $\mathbb{R}^3$ the rigid body motions are translations along $x$, $y$, $z$ and rotations about $x$, $y$, $z$.

Finally, the concept of \textit{minimum rigidity} is introduce. 
\begin{def3}  \citep{Anderson2008}.
A graph is minimally rigid if it is rigid and the removal of a single edge causes it to lose rigidity. Mathematically, this condition can be checked by the number of edges $N_e$, so that if $N_e=2N-3$ in $\mathbb{R}^2$ or $N_e=3N-6$ in $\mathbb{R}^3$ the graph is minimally rigid.
\end{def3}

\subsection{Delaunay Triangulation} \label{sec:DelaunayTriangulation}

The following definitions and concepts are the basics for 2D Delaunay triangulations.
\begin{def4}
A triangulation of a set $\mathcal{P}$ points is a planar graph with vertices at the coordinates $p_i \in \mathcal{P}$ and edges that subdivide the convex
hull $H(\mathcal{P})$ into triangles, so that the union of all triangles equals the convex hull.
\end{def4}

Any triangulation with $N$ vertices consists of $2(N-1)-N_b$ triangles
and has $N_e = 3 (N-1) -N_b$ edges, where $N_b$ denotes the number of agents on the boundary $\partial H(\mathcal{P})$ of
the convex hull. The edges of a triangulation do not cross each other. Furthermore, the triangulation of $N > 3$ points is not unique. The Delaunay
triangulation is a proximity graph that can be constructed by the geometrical configuration of the vertices.

\begin{def5}\label{def:incircleTest}\citep{Delaunay1934}. A triangle of a given triangulation of a set $\mathcal{P}$ of points is said to be Delaunay if there is no point $p_i\in\mathcal{P}$ in the interior of its circumcircle. 
\end{def5}

The circumcircle of a triangle is the unique circle passing through its three vertices.

\begin{def6} \citep{Delaunay1934}. A Delaunay triangulation is a triangulation in which all triangles satisfy the local Delaunay property.
\end{def6}

\begin{exp2}
Figure \ref{fig:delaunayTriang} shows an example of the possible triangulations for the set of points $\{A,B,C,D,E\}$. Only the one on the left is a Delaunay triangulation. In the middle, point $C$ is in the interior of the circumcircle of the triangle formed by $ABE$. On the right, points $B$ and $C$ lie inside the circumcircle of the triangle formed by $ADE$.
\begin{figure}[ht!]
    \centering
    \includegraphics[width=0.95\hsize]{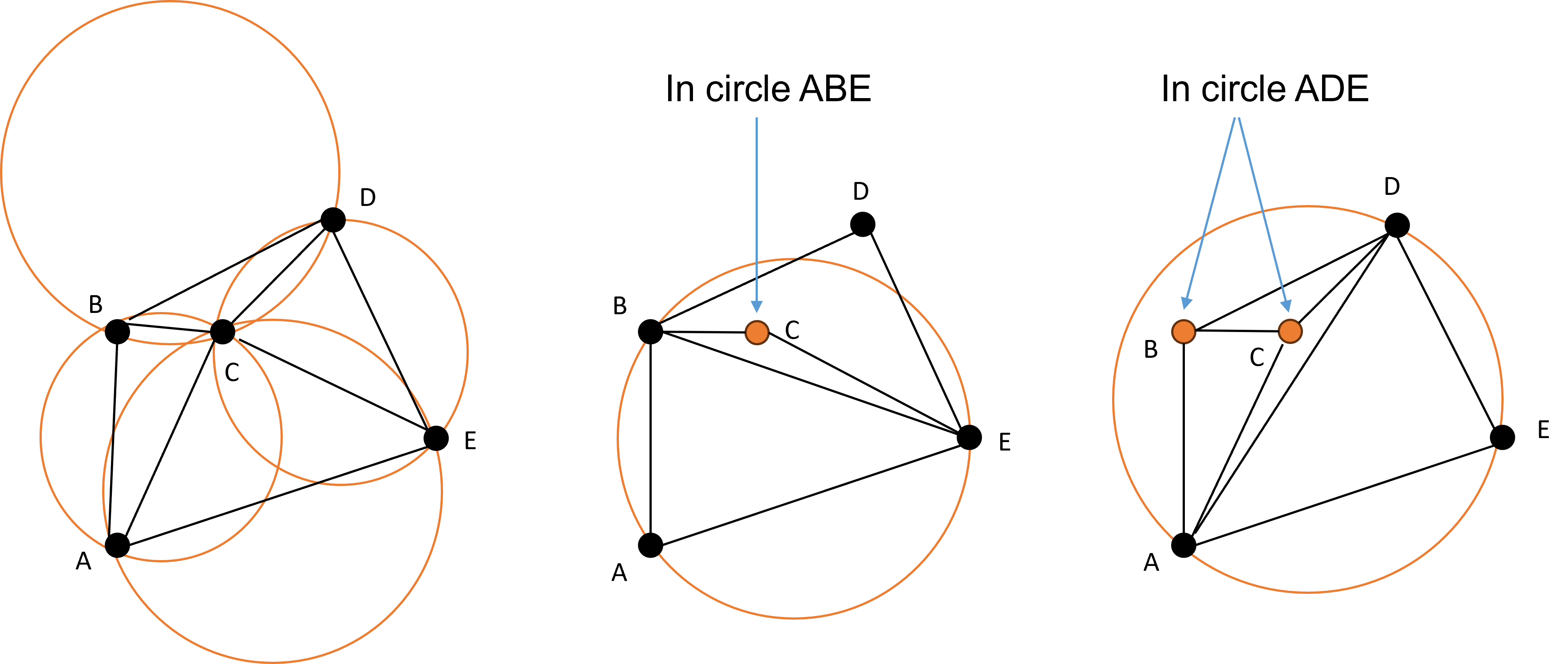}
    \caption{Example of the possible triangulations for the set of points $\{A,B,C,D,E\}$. Only the one on the left is a Delaunay triangulation.}\label{fig:delaunayTriang}
\end{figure}
\end{exp2}

Similar definitions follow for 3D triangulations, where the convex hull of $\mathcal{P}$ is
decomposed into tetrahedra such that the vertices of tetrahedra belong to $\mathcal{P}$, and the intersection of two tetrahedra is either empty or a vertex or an edge or a face. For such a reason,  a triangulation in 3D space can be called triangulation, 3D triangulation,
or {\it tetrahedralization} \citep{Toth2017}.

A framework whose graph $\mathcal{G}$ is a Delaunay triangulation is rigid and the rank of the rigidity matrix is $2N-3$ (respectively $3N-6$) in $\mathbb{R}^2$ (respectively $\mathbb{R}^3$) \citep{Eren2002}.

\section{Problem description} \label{sec:problemDescription}
\subsection{Agents model}
The state of each mobile agent is described by the vector
\begin{equation}
    p_i(t)=\begin{pmatrix}
    p_{x,i}(t) \\
    p_{y,i}(t) \\
    p_{z,i}(t) \end{pmatrix},
\end{equation}
which represents the Cartesian coordinates.

Let the $N$ agents obey the single-integrator dynamics:
\begin{equation} \label{eq:sysDynamics}
	\dot{p}_i(t)=u_i(t), \ \ i=1,\dots,N,
\end{equation}
where $u_i(t) \in \mathbb{R}^3$ are the control inputs of agent $i$, which will be described later in the paper.

We assume that each agent is equipped, at least, with hardware that allows the measurement of the distance to other agents and relative position measurements in their local coordinate
frames. 

\subsection{Gradient control}
In \citet{Krick2009}, a distributed control law is proposed for formation control, where the control law is derived from a potential function based on an undirected and infinitesimally rigid graph. More specifically, the potential function has the form
\begin{equation} \label{eq:potentialKrick}
    W=\frac{1}{4}\sum_{(i,j)\in\mathcal{E}}(d^2_{ij}-{d^*_{ij}}^2)^2,
\end{equation}
where $d_{ij}=\|p_i-p_j\|$ and $d^*_{ij}$ is the prescribed distance for the edge $(i,j)\in\mathcal{E}$.
The gradient descent control law for each agent $i$ derived from the potential function \eqref{eq:potentialKrick} is then
\begin{equation} \label{eq:controlKrick}
    u_i=-\nabla_{p_i}W=-\sum_{j\in\mathcal{N}_i}(d^2 _{ij}-{d^*_{ij}}^2)(p_i-p_j).
\end{equation}
It has been shown in \citep{Krick2009} that, for a single integrator model of the agents moving in $\mathbb{R}^2$, the target formation is \textit{local asymptotically stable} under the control law \eqref{eq:controlKrick} if the graph of the framework is infinitesimally rigid. However, the global stability analysis
beyond a local convergence for formation control systems with general shapes cannot be achieved due to the existence of multiple equilibrium sets, and a complete analysis of these sets and their stability property is very challenging due to the nonlinear control terms \citep{Sun2015}. More specifically, even though $W=0$ in \eqref{eq:potentialKrick} only at the desired formation, i.e., when $d_{ij}=d_{ij}^*$, there exist other equilibria sets that correspond to $\nabla_{p_i}W=0$, including collinearity  (in $\mathbb{R}^2$) and collinearity and coplanarity (in $\mathbb{R}^3$) of the agents. 

\subsection{Shield model} \label{sec:shieldmodel}
The team of agents should be deployed to protect a certain area of interest that, without loss of generality, is placed around the origin, i.e., $p_0^*=\mathbf{0}$. For the aforementioned purpose, the agents form a mesh with a certain shape that we call a \textit{shield}. 
We model this ``virtual'' shield by a quadric surface $\mathcal{S}\in \mathbb{R}^3$ described in the following compact form
\begin{equation}
\label{eq:shieldSurface}
    \mathcal{S}\equiv p^\top Q_1p+Q_2=0,
\end{equation}
where $p\in\mathbb{R}^3$, $Q_1 \in \mathbb{R}^{3\times 3}$ such that $Q_1=Q_1^\top$, and $Q_2\in\mathbb{R}$. Note that this is a quite general form though it excludes some shapes such as the different paraboloids or the parabolic cylinder.
Additionally, since the shield is deployed around the point $p_0^*=\mathbf{0}$, we consider quadric surfaces in their normal form \citep{Venit1995}, which imposes some constraints on the values of $Q_i$. Additionally, we also exclude the case of non-connected surfaces, i.e., from any point of the surface, a continuous path can be drawn to any other point of it without crossing its boundary.

\begin{figure}[th]
  \centering
  \parbox{\figrasterwd}{
    \parbox{.32\figrasterwd}{%
      {\includegraphics[width=\hsize]{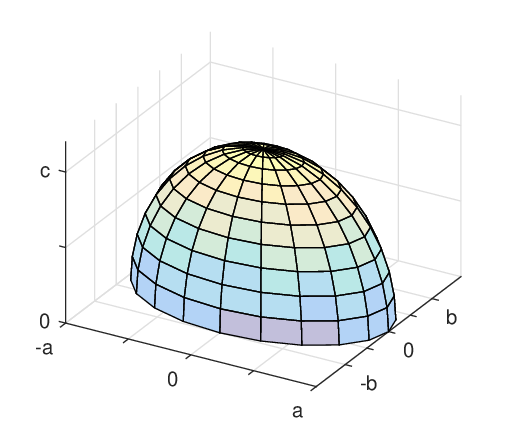}}
    }
    \parbox{.32\figrasterwd}{%
      {\includegraphics[width=\hsize]{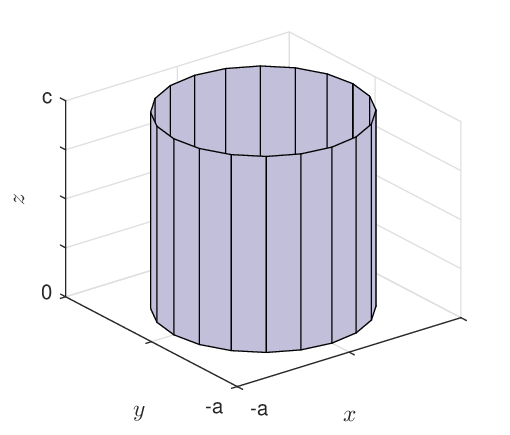}}
    }
    \parbox{.32\figrasterwd}{%
      {\includegraphics[width=\hsize]{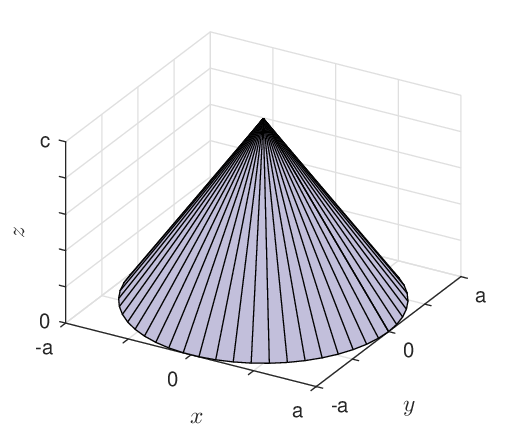}}
    }
    }
  \caption{View of the shield examples given in Table~\ref{tab:shieldexample}: Semi-ellipsoid (left), cylinder (middle), cone (right).}
  \label{fig:shieldexample}
\end{figure}

Furthermore, the shield might require the definition of some additional constraints for the positioning of the agents, for example, having an upper and/or lower bound on some of the coordinates, but this will be handled by the control law. In general, we constrain $z>0$. Table~\ref{tab:shieldexample} and Figure~\ref{fig:shieldexample} illustrate some examples of $\mathcal{S}$. 
\begin{table}[th]  
\renewcommand{\arraystretch}{1.5}
\centering
\begin{tabular}{|m{9em}|c|c|c|} 
 \hline
 \textbf{Type of shield} & $\mathbf{Q_1}$ & $\mathbf{Q_2}$ &  \textbf{Constraints} \\ 
 \hline
 \hline
 Semi-ellipsoid & $\begin{pmatrix}\frac{1}{a^2} & 0 & 0 \\ 0 & \frac{1}{b^2}  & 0 \\ 0 & 0 & \frac{1}{c^2} \end{pmatrix}$ & $-1$ & $z\geq 0$ \\ 
 \hline
 Cylinder of height $c$ & $\begin{pmatrix}\frac{1}{a^2} & 0 & 0 \\ 0 & \frac{1}{a^2}  & 0 \\ 0 & 0 & 0\end{pmatrix}$ &  $-1$ & $c\geq z\geq 0$ \\ 
 \hline
  Cone of height $c$ & $\begin{pmatrix}\frac{1}{a^2} & 0 & 0 \\ 0 & \frac{1}{a^2}  & 0 \\ 0 & 0 & -\frac{1}{c^2}\end{pmatrix}$ & $-1$ & $c\geq z\geq 0$ \\ 
 \hline
\end{tabular}
\caption{Examples of shield models given by \eqref{eq:shieldSurface}.}
\label{tab:shieldexample} 
\end{table}

For the state of an agent $i$, we can define the following function
\begin{equation} \label{eq:fsurface}
f_\mathcal{S}(p_i)=p_i^\top Q_1p_i+Q_2.
\end{equation}
Note that $f_\mathcal{S}(p_i)=0 \iff p_i\in\mathcal{S}$.

\subsection{Problem statement}

We can announce the problem as follows:
\\
\textbf{Problem 1.} \textit{Given the team of agents \eqref{eq:sysDynamics} whose topology is modeled by a graph $\mathcal{G}=(\mathcal{V},\mathcal{E})$ and the virtual shield described by \eqref{eq:shieldSurface}: I) Design an algorithm that finds the set of edges $\mathcal{E}$ and the corresponding set of target distances for the formation control, $\{d_{ij}^*: \ (i,j)\in\mathcal{E}\}$, such that the team is deployed forming the shield in an almost uniform distribution over the surface; II) Design the distributed control law $u_i(t)$ for each agent $i$
\begin{equation}
    u_i(t)=f_i(p_i,\{p_i-p_j, \ d_{ij}^*, \ j\in\mathcal{N}_i\},f_\mathcal{S}), 
\end{equation}
\textit{such that for each neighboring node $j\in\mathcal{N}_i$, the Euclidean distance between them, $d_{ij}=\|p_i-p_j\|$, satisfies}
\begin{equation}
    \lim_{t\to\infty} d_{ij}(t)=d_{ij}^*, \ j\in\mathcal{N}_i,
\end{equation}
\textit{while lying on the virtual surface}
\begin{equation}
    \lim_{t\to\infty} f_\mathcal{S}(p_i)=0.
\end{equation}}

Therefore, the solution to this problem is developed in two steps. The first one (corresponding to the first objective) is presented in Section \ref{sec:ShieldBuilding}; and Section \ref{sec:controllaw} provides the proposed solution to the second objective.

\section{Shield building} \label{sec:ShieldBuilding}
\begin{figure}[ht!]
    \centering
    \includegraphics[width=0.8\hsize]{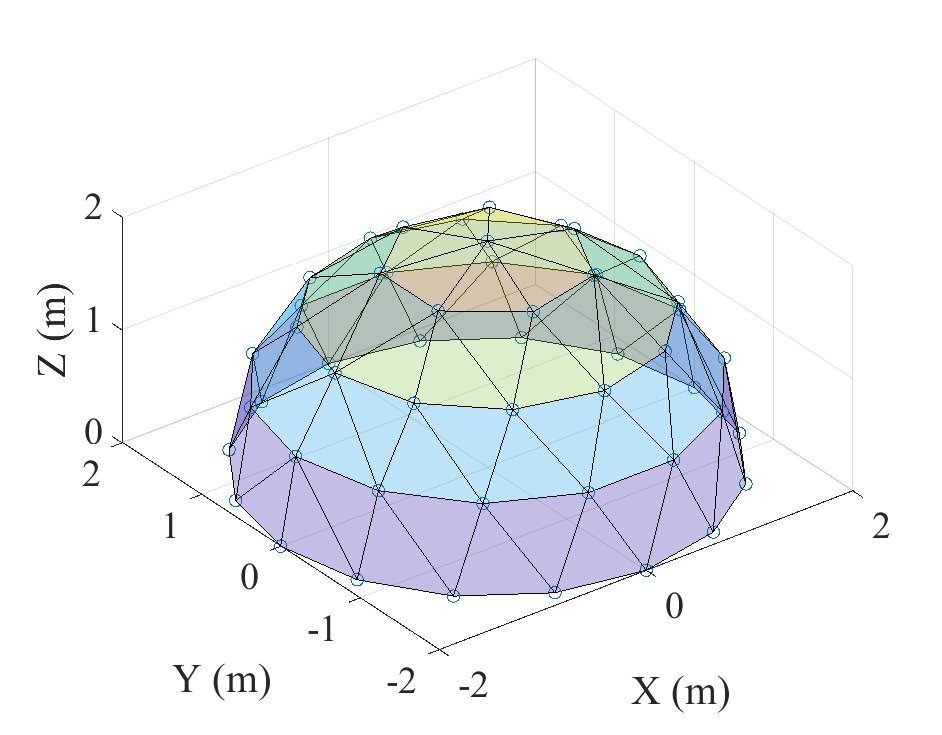}
    \caption{Target formation example. A team of $N=50$ agents forms a semi-spherical shield.}\label{fig:exampleShield}
\end{figure}

This section presents a method to design the target formation that consists of a mesh of nodes forming the shield. An example of a shield in which the virtual surface is a semi-sphere is shown in Figure \ref{fig:exampleShield}. 

First, an algorithm is proposed so that, given a desired shape, its geometry, and the number of agents, an estimation of the formation's target distances is given such that the agents distribute more or less uniformly over the virtual surface. After that, a procedure to create the links between nodes is presented so that the result is a Delaunay triangulation. 

There exist in the literature many results that study how to distribute points over a sphere. The foundation of this is the so-called Thomson problem \citep{Tomson1904}: \textit{find the minimum electrostatic potential energy configuration of $N$ electrons constrained on the surface of the unit sphere}, and it is being around for more than a century. This problem seems simple in its formulation, but it is one of the mathematical open problems due to the complexity of the general solution, and the computability or tractability of some simple cases. Thus, there exist solutions based on numerical analysis and approximation theory such as: Fibonacci and generalized spiral nodes; projections of low discrepancy nodes from the unit square; polygonal nodes such as icosahedral, cubed sphere, and octahedral nodes; minimal energy nodes; maximal determinant nodes; or random nodes (see \citep{Hardin2016,Koay2011} and references therein). However, the extrapolation to other surfaces is not straightforward and requires complex mathematics that ends in different approximations \citep{Kreyszig2007}.

Thus, the proposed method tries to find a simple procedure to provide an initial estimation of the maximum distance between agents that allows the placement of the nodes over the surface. It is based on the idea that the area of the surface, $A_\mathcal{S}$, is a generally well-known property that only depends on a few parameters. The idea behind this consists of approximating the area of the surface by the area of Delaunay triangles of the formation, assuming than are equilateral, to infer the distance between nodes.

For an equilateral triangle, if the distance between the points is $d$ and $h$ its height, the area is given by
\begin{equation}
    A_f=\frac{d\cdot h}{2}=\tfrac{\sqrt{3}}{4}d^2.
\end{equation}
According to the theory of rigid formations \citep{Gallier2011} the number of triangles of a Delaunay triangulation in 2D is given by
\begin{equation} \label{eq:nfaces}
    f=2N-2-e_b,
\end{equation}
where $e_b$ is the number of edges in the boundary of the triangulation. Note that even though the state space is $\mathbb{R}^3$, the fact that the target formation is constrained over $\mathcal{S}$, makes the previous result applies. 

Thus, the area of the set of triangles is
\begin{equation} \label{eq:areaTriangulation}
    f\cdot A_f=(2N-2-e_b)\tfrac{\sqrt{3}}{4}d^2.
\end{equation}
On the other hand, the number of edges in the boundary also depends on geometrical properties of the surface. For instance, if the boundary is defined by the intersection of the surface with a plane, the result is a curve whose length can be approximated by the number of nodes in the curve and the distance between them, i.e., $L_b\approx e_b\cdot d$. Actually, this is the perimeter of the boundary of the triangulation. This yields in \eqref{eq:areaTriangulation} to
\begin{equation} \label{eq:areaTriangulation2}
    f\cdot A_f\approx (2N-2-\frac{L_b}{d})\tfrac{\sqrt{3}}{4}d^2.
\end{equation}
If the area of the surface $A_{\mathcal{S}}$ is approximated by \eqref{eq:areaTriangulation2}, this results on a second order equation to solve $d$:
\begin{equation} \label{eq:areaApproximation}
   A_{\mathcal{S}}\approx (2N-2-\frac{L_b}{d})\tfrac{\sqrt{3}}{4}d^2.
\end{equation}
For a convex surface, \eqref{eq:areaApproximation} is actually an inequality $A_{\mathcal{S}}\geq  (2N-2-\frac{L_b}{d})\tfrac{\sqrt{3}}{4}d^2$, so that
\begin{equation} \label{eq:dbound}
    d \leq \frac{L_b+\sqrt{L_b^2+\tfrac{32}{\sqrt{3}}A_\mathcal{S}(N-1)}}{4(N-1)}.
\end{equation}

Note that the previous procedure not only provides a value for the maximum inter-distance between nodes $d$ but the number of nodes that should be placed in the boundary, since the number of vertices of a closed path or a cycle equals the number of edges. 
\begin{equation} \label{eq:nb}
    n_b=e_b=\lceil\frac{L_b}{d}\rceil,
\end{equation}
where $\lceil x \rceil$ is the ceiling function. 
If we assume that the nodes are distributed on the surface in rings of different heights, the previous procedure can be repeated iteratively to determine the height and the number of nodes in each ring. The idea is as follows. Let us denote $h_k$ the height of the ring $k$, $A_{k}$ the area of the resulting surface over the intersection of $\mathcal{S}$ with plane $z=h_k$, $L_k$ the perimeter of such plane section and $N_k$ the remaining number of nodes at iteration $k$. Then, if $A_k$ and $L_k$ can be expressed in terms of $h_k$ and $d$ is given by \eqref{eq:dbound}, then an equivalent equation to \eqref{eq:areaApproximation} can be applied to get $h_k$:
\begin{equation} \label{eq:hk}
      A_k(h_k)\approx (2N_k-2-\frac{L_k(h_k)}{d})\tfrac{\sqrt{3}}{4}d^2,
\end{equation}
where $N_k$ is the number of remaining nodes. Thus, the number of nodes to be placed at the ring of height $h_k$ is
\begin{equation} \label{eq:nk}
    n_k=\lceil\frac{L_k}{d}\rceil.
\end{equation}
\begin{rmk5}
The ceiling operation in \eqref{eq:nk} makes that, in general, $n_k\cdot d > L_k$. Then, the parameter $d$ can be adjusted for each level $k$ as 
\begin{equation} \label{eq:dk}
    d_k=\frac{L_k}{n_k},
\end{equation}
so that all the agents can be uniformly distributed in the ring of heigh $h_k$. Moreover, when the algorithm is in the last step, it might occur that the number of remaining agents, $N_k$, satisfies that $N_k<n_k$. In that case, $n_k$ is set to $N_k$, and then $d_k$ is computed by \eqref{eq:dk}. That way, the algorithm always guarantees a position for each node.
\end{rmk5}
Algorithm \ref{alg:shieldbuilding} summarizes the iterative procedure for building the shield. As input parameters, it receives the number of nodes and some parameters of the surface $\mathcal{S}$ such as the area $A_\mathcal{S}$, the length of the boundary $L_b$, and $h_{max}$, which is the height of the surface. This parameter can be given implicitly in some quadrics (ellipsoid, sphere) or might be specified in other cases such as some of the examples presented in Table \ref{tab:shieldexample} and Figure \ref{fig:exampleShield}. Moreover, the number of nodes is bounded as $N\geq4$, i.e., the minimal configuration is a tetrahedron. As a result, it returns a set of triples $\{n_k, d_k, h_k\}$ with $k=0,\dots,K-1$, where $K\geq 2$ is the number of rings. 
\begin{algorithm}
\caption{Algorithm for shield building}\label{alg:shieldbuilding}
\begin{algorithmic}
\STATE
\STATE \textbf{Input: }$\ \ N, A_{\mathcal{S}}, L_b, h_{max}$
\STATE \textbf{Output: }$\{\{n_k, d_k, h_k\}, \ k=0,\dots,K-1\}$
\STATE \hspace{0.5cm} Compute $d$ as \eqref{eq:dbound}
\STATE \hspace{0.5cm} Compute  $n_b$ as \eqref{eq:nb}
\STATE \hspace{0.5cm} Adjust  $d_0$ according to $d_0=\frac{L_b}{n_b}$ 
\STATE \hspace{0.5cm} $k \gets 1$
\STATE \hspace{0.5cm} $N_k\gets N-n_b$
\STATE \hspace{0.5cm} Compute  $h_k$ as the solution of \eqref{eq:hk}
\STATE \hspace{0.5cm} \textbf{while } $N_k > 0$ and $h_k\leq h_{max}$
    \STATE \hspace{1cm} Compute  $n_k$ as \eqref{eq:nk}
    \STATE \hspace{1cm} \textbf{if }  $n_k>N_k$
    \STATE \hspace{1.5cm} $n_k\gets N_k$
    \STATE \hspace{1cm} \textbf{end if}
    \STATE \hspace{1cm} Compute  $d_k$ as \eqref{eq:dk}
    \STATE \hspace{1cm} $k \gets k+1$
    \STATE \hspace{1cm} $N_k\gets N_k-n_k$
    \STATE \hspace{1cm} Compute $h_k$ as the solution of \eqref{eq:hk}
\STATE \hspace{0.5cm} \textbf{end while }
\end{algorithmic}
\end{algorithm}

\begin{exp1}Let us consider a semi-sphere of $R$=15. Let us compute the solution provided by the proposed method and estimate the error of the estimation. Table \ref{tab:algExample}  shows the estimation for the distance between nodes $d$ for different values of $N$ and the error in the estimated area of the surface. The number of triangles and the number of nodes in the boundary are also given.

\begin{table}[th]  
\renewcommand{\arraystretch}{1.3}
\centering
\begin{tabular}{||c c c c c||} 
 \hline
 $N$ & $d$ & $f$ & $e_b$ & $(A_\mathcal{S}-fA_f)/A_\mathcal{S}$ \\ [0.5ex] 
 \hline\hline
 20 &	10.59 &	29 &	9	& 0.29 \\ 
 \hline
 50 & 6.27 & 82 & 16	& 0.01 \\
 \hline
 100	& 4.31 &  176 & 22 & $6.28\cdot 10^{-4}$ \\ 
 \hline
\end{tabular}
\caption{Values obtained for $d$ in Algorithm \ref{alg:shieldbuilding} for a semi-sphere of $R=15$ for different values of $N$.}
\label{tab:algExample} 
\end{table}

The results show that the larger the value of $N$, the better the approximation and, of course, the shorter the distance between nodes. 
\end{exp1}

Once the number of agents that should be placed in each level $h_k$ is computed, a simple procedure to create the edges can be followed as follows:
\begin{itemize}
    \item Each point $i$ creates a link to the two adjacent points in the ring of height $h_k$: ``left'' ($i-1$) and ``right'' ($i+1$).
    \item Each point $i$ of the level $h_k$ creates a link to the points $j$ of the level $h_{k+1}$ that are at a distance $d_{ij}\leq d+\varepsilon$, where $d$ is computed by means of \eqref{eq:dbound} and $\varepsilon$ is a design parameter.
    \item If the projection over $z=0$ of the new link $ij$ between levels $h_k$ and $h_{k+1}$ intersects with the projection of an existing link between these levels, the link $ij$ is removed.
    \item Update $d_{ij}^*$ to the actual value $d_{ij}$.
\end{itemize}
The previous method creates a triangulation where, in general, the triangles are not equilateral as it was assumed, at first, when computing the approximate value $d$. Hence, the target values $d_{ij}^*$ will be different from the initial estimation. We first assume that the resulting triangulation is Delaunay. Section \ref{sec:Delaunay} will provide a method to check this condition for each triangle.

\begin{rmk8}
A conservative value for $\epsilon$ can be defined by noting that the separation between two consecutive rings can also be bounded by $d$. Since $d_k\leq d$ holds in \eqref{eq:dk}, if such upper bound is an equality, the Delaunay condition imposes that the distance $d_{ij}$ is bounded by $d_{ij}^2\leq 2d^2$ (rectangle triangle). Then, an upper bound for $\epsilon$ is $\epsilon\leq (\sqrt{2}-1)d$.
\end{rmk8}

The following result estimates the upper and lower bounds for the number of edges in a triangulation generated by the procedure described in this section. 

\begin{prop1} \label{prop:prop1}
Let us consider a network of $N$ nodes deploying a formation in form of a Delaunay triangulation over a surface $\mathcal{S}$. The number of edges of the triangulation $N_e$ is bounded as
\begin{equation}
    2N-2\leq N_e\leq 3N-6. 
\end{equation}
\end{prop1}
\begin{proof} Similarly to \eqref{eq:nfaces}, the number of edges is also a linear function of the number of vertices and boundary edges \citep{Gallier2011}. More specifically, and according to the \textit{Euler Formula}, it holds that
$$N-e_b-e_i+f=1,$$
where $e_i$ are the number of edges not in the boundary. Also, it holds that $3f=e_b+2e_i$ since each non-boundary edge is shared by two faces, and then it follows that $e_i = 3N-3-2e_b$. 

The minimal configuration of the shield requires at least 3 nodes in the boundary so that a valid triangulation is generated (and the minimum number of nodes is $N=4$). Then, the total number of edges can be bounded as
$$N_e=e_i+e_b=3N-3-2e_b+e_b=3N-3-e_b\leq 3N-6.$$
Similarly, the maximum $e_b$ is $N-1$, which corresponds to having $N-1$ nodes in the boundary. Then
$$N_e=e_i+e_b=3N-3-e_b\geq 2N-2,$$
which completes the proof.
\end{proof}

\subsection{Local Characterization of Delaunay Triangulations} \label{sec:Delaunay}
In this section, we present a method to check if the formation of agents in form of triangulation deployed in a surface $\mathcal{S}$ \eqref{eq:shieldSurface} is Delaunay's. The basic definitions were introduced in Section \ref{sec:DelaunayTriangulation}. Each of the vertices of the triangulation represents one agent $i\in\mathcal{V}$ with the coordinates $p_i\in\mathbb{R}^3$. As shown in \citet{Mathieson2019}, if the connectivity graph $\mathcal{G}$ is a Delaunay triangulation, each agent $i$ is connected to its geometrically closest neighbors. 

\begin{figure}[th]
\centering
\includegraphics[width=0.3\linewidth]{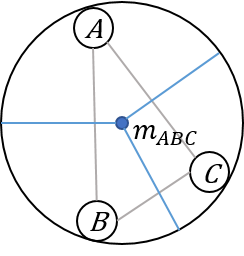}
\caption{The triangle formed by $\{A,B,C\}$, its circumcircle, and its circumcenter $m_{ABC}$.}
\label{fig:DelaunayTri}
\end{figure}

The next analysis will provide a local characterization so that each agent $i$ can check if a triangle is Delaunay by exploiting the empty-circumcircle property of Definition \ref{def:incircleTest}. We particularly extend the ideas of \citet{Schwab2021} which deal with proximity graphs in $\mathbb{R}^2$ to a surface $\mathcal{S}\in\mathbb{R}^3$ defined by \eqref{eq:shieldSurface}. In Figure \ref{fig:DelaunayTri} a 2D view of a triangle and its circumcircle is depicted to illustrate the concepts. The point $m_{ABC}$ represent the circumcenter of the triangle formed by the three vertices $\{A,B,C\}$. 

Consider the following matrix
\begin{equation}
    O_{ABC}=\begin{pmatrix}p_{x,A} & p_{x,B} & p_{x,C} \\
                           p_{y,A} & p_{y,B} & p_{y,C} \\
                           p_{z,A} & p_{z,B} & p_{z,C}\end{pmatrix},
\end{equation}  
denoted as the \textit{orientation matrix}. Note that $|O_{ABC}|=0$ if the three points $A$, $B$, and $C$ are collinear. Also, the three points define a plane in the space $\mathbb{R}^3$:
\begin{equation} \label{eq:planeDef}
    f_{\pi}\equiv ap_x+bp_y+cp_z+d=0,
\end{equation}
where $a, b, c, d \in \mathbb{R}$ can actually be related to the coordinates of $p_{BA}=p_B-p_A$ and $p_{CA}=p_C-p_A$ as
\begin{align}
    a=&\begin{vmatrix}p_{y,BA} & p_{y,CA} \\ p_{z,BA} & p_{z,CA}\end{vmatrix}, \
    b=-\begin{vmatrix}p_{x,BA} & p_{x,CA}\\ p_{z,BA} & p_{z,CA}\end{vmatrix}, \label{eq:ab_plane}\\
    c=&\begin{vmatrix}p_{x,BA} & p_{x,CA} \\ p_{y,BA} & p_{y,CA}\end{vmatrix}, \
    d=-|O_{ABC}|. \label{eq:cd_plane}
\end{align}  

\begin{rmk3}Note that if $p_A, \ p_B$, and $p_C$ are collinear or if the origin $(0, 0, 0) \in f_\pi$ in \eqref{eq:planeDef}, then $|O_{ABC}|=0$. However, these situations cannot occur when the distribution of points and connection between them is generated as explained at the beginning of this section. First, three points in a non-degenerate quadric surface cannot be collinear; secondly, in the case of degenerate quadrics, the alignment of three points forming a triangle is excluded in the procedure to generate the formation; and finally, the shield is deployed around the area of interest, centered at the origin, and there are not three points in the plane $p_z=0$ forming a triangle.
\end{rmk3}

\subsubsection{Circumcenter}
The circumcenter of the triangle defined by the points $p_A, p_B$, and $p_C$ is denoted by $m_{ABC}$ and satisfies the following condition:
\begin{equation} \label{eq:CC1}
    \|m_{ABC}-p_A\|=\|m_{ABC}-p_B\|=\|m_{ABC}-p_C\|,
\end{equation}
i.e., the distance from each vertex is the same. However, in $\mathbb{R}^3$ there are infinite points that fulfill such conditions. Therefore, the following constraint is required to compute the circumcenter: $m_{ABC}\in f_\pi$. 

Then following result provides a method to compute it as well as the radius of the circumcircle.

\begin{lem4} \label{lem:lem1}
The circumcenter $m_{ABC}$ of three points $p_A$, $p_B$, and $p_C \in\mathbb{R}^3$ that are not collinear is given by 
\begin{equation} \label{eq:lemma11}
    \begin{pmatrix}m_{ABC} \\ \gamma\end{pmatrix}
    = \frac{1}{2}\Lambda^{-1}\begin{pmatrix}\|p_A\|^2 \\ \|p_B\|^2 \\ \|p_C\|^2 \\ 2|O_{ABC}|\end{pmatrix},
\end{equation}
where 
\begin{equation} \label{eq:Lambda}
    \Lambda=\begin{pmatrix}O_{ABC} & \mathbf{1} \\
    v^\top & 0\end{pmatrix},
\end{equation}
$\mathbf{1}=(1 \ 1 \ 1)^\top$, $\gamma$ is a scalar, and $v$ is the normal vector of the plane \eqref{eq:planeDef} defined by $p_A$, $p_B$ and $p_C$. Furthermore, the radius of the circumcircle $r_{ABC}$ is
\begin{equation} \label{eq:lemma12}
    r_{ABC}=\sqrt{2\gamma+\|m_{ABC}\|^2}.
\end{equation}
\end{lem4}

\begin{proof}
The proof can be found in the Appendix.
\end{proof}

\begin{lem5}
If $p_A$, $p_B$, and $p_C$ are not collinear and the origin $(0,0,0)\notin f_\pi$ defined in \eqref{eq:planeDef}, the determinant of the matrix $\Lambda$ in \eqref{eq:Lambda} is always negative, i.e., $|\Lambda|<0$.
\end{lem5}

\begin{proof}
The determinant of $\Lambda$ in \eqref{eq:Lambda} can be computed following the  Laplace expansion as
\begin{align*}
    |\Lambda|=&-a\begin{vmatrix}p_{y,A} & p_{z,A} & 1 \\ p_{y,B} & p_{z,B} & 1 \\ p_{y,C} & p_{z,C} & 1\end{vmatrix}+b\begin{vmatrix}p_{x,A} & p_{z,A} & 1 \\ p_{x,B} & p_{z,B} & 1 \\ p_{x,C} & p_{z,C} & 1\end{vmatrix} \\
&-c\begin{vmatrix}p_{x,A} & p_{y,A} & 1 \\ p_{x,B} & p_{y,B} & 1 \\ p_{x,C} & p_{y,C} & 1\end{vmatrix}.
\end{align*}

As the determinants above are another way of computing the parameters $a$, $-b$, and $c$, respectively, of the plane $f_\pi$ \citep{Anton2013}, then it follows that
$$|\Lambda|=-a^2-b^2-c^2<0.$$
\end{proof}

\subsubsection{In-spherical cap test}
To test if a triangle formed by three points $p_A$, $p_B$, and $p_C$ is Delaunay locally, we can check that no other node of the network, represented by $p_D$, satisfies that $\|p_D-m_{ABC}\|<r_{ABC}$. This is an extension of the incircle test \citep{Schwab2021} to our setting $\mathcal{S}\in\mathbb{R}^3$ \eqref{eq:shieldSurface}. For that purpose, we define the following matrix
\begin{equation} \label{eq:Mabcd}
    M_{ABCD}=\begin{pmatrix}\Lambda & v_P \\
        \begin{pmatrix}p_D^\top & 1\end{pmatrix} & \|p_D\|^2\end{pmatrix},
\end{equation}
where 
\begin{equation} \label{eq:vp}
    v_P^\top=(\|p_A\|^2 \ \|p_B\|^2 \ \|p_C\|^2 \ 2|O_{ABC}|).
\end{equation}

Note that $|M_{ABCD}|=0$ if $\|p_D-m_{ABC}\|=r_{ABC}$, since if we define $q^\top=(-2x_{ABC} \ -2y_{ABC} \ -2z_{ABC} \ \|m_{ABC}\|^2 \ 1)\neq \mathbf{0}$, it holds that $M_{ABCD}\cdot q=\mathbf{0}$.

The next result shows that studying the sign of $|M_{ABCD}|$ allows to determine if a candidate point $p_D$ of the mesh breaks or not the condition that the triangle formed by $p_A$, $p_B$, and $p_C$ is Delaunay.

\begin{thm3} \label{thm:delaunay}
Consider three non-collinear points $p_A$, $p_B$, and $p_C$ forming a plane $f_\pi$ defined in \eqref{eq:planeDef} such that $(0,0,0)\notin f_\pi$, and a candidate point $p_D$.
Let $m_{ABC}$ the circumcenter of $p_A$, $p_B$, $p_C$ computed as the solution of \eqref{eq:lemma11} and $r_{ABC}$ the radius \eqref{eq:lemma12} of the circumcircle.  The point $p_D$ satisfies $\|p_D-m_{ABC}\|<r_{ABC}$ if $|M_{ABCD}|>0$.
\end{thm3}
\begin{proof}
The proof can be found in the Appendix.
\end{proof} 

\begin{rmk6}
The previous result would allow to change the topology of the system dynamically if we want the condition given in Theorem \ref{thm:delaunay} to be satisfied at any time while the agents are moving, but this is out of the scope of the paper. Switching topologies will be part of the future work.
\end{rmk6}
\begin{rmk7}
It must be noticed that the Delaunay extension presented in the paper is not a true 3D extension because it is not based on tetrahedrons forming the volume under the quadratic surface. The new method could be labeled as a \textit{2D+ extension}.
\end{rmk7}

\section{Control law} \label{sec:controllaw}

Consider the following potential function:
\begin{equation} \label{eq:potential}
    W=\frac{\kappa_1}{4} \sum_{(i,j)\in\mathcal{G}}(d_{ij}^2-d_{ij}^{*2})^2+\frac{\kappa_2}{4}\sum_{i=1}^N(f_{\mathcal{S}}(p_i))^2,
\end{equation}
where $d_{ij}=\|p_i-p_j\|$ is the distance between two agents $i$ and $j$, $d_{ij}^{*}$ is the prescribed inter-distance between both agents in the objective formation, and $\kappa_1, \kappa_2\in\mathbb{R}_{>0}$. Note that \eqref{eq:potential} includes a first term corresponding to \eqref{eq:potentialKrick} and an additional term that takes into account how far each agent is from the surface $\mathcal{S}$.

Then, the distributed control law to achieve the desired objective can be computed as
\begin{equation} \label{eq:controlLaw}
    u_i=-\frac{\partial W}{\partial p_i}=-\kappa_1\sum_{j\in\mathcal{N}_i}(d_{ij}^2-d_{ij}^{*2})(p_i-p_j)-\frac{\kappa_2}{2}f_{\mathcal{S}}(p_i)\frac{\partial f_{\mathcal{S}}(p_i)}{\partial p_i}.
\end{equation}

\begin{rmk1} At a first stage, we do not consider the constraints of the surface $\mathcal{S}$ on $z$ since we are interested on studying the analytical properties of the proposed control law. Then, we will introduce a modified control law to consider such constraints.
\end{rmk1}

\begin{rmk9}
The feedback gains $\kappa_1$ and $\kappa_2$ should be chosen in such a way that both terms contribute in a similar scale. Note that the quadric surface is defined in normal form, so the evaluation of $f_\mathcal{S}(p_i)$ is in the scale of 1 Additionally, its gradient is somehow normalized since $p_i$ is weighted by $Q_1$. By contrast, the term of the formation shape control depends on the square of distances and it is a summation in the set of neighbors. Thus, a choice of $\kappa_2\sim\frac{\bar{|\mathcal{N}_i|}d^2}{\|Q_1\|}\kappa_1$, where $\bar{|\mathcal{N}_i|}$ is the average of neighboring nodes and $\|Q_1\|$ the matrix norm of $Q_1$, is a fair approximation. Alternatively, an upper bound for $\bar{|\mathcal{N}_i|}$ can be taken. Note that with the generated topology, setting an upper bound for this number is easy.
\end{rmk9}

\begin{rmk11}
The outcomes of Algorithm \ref{alg:shieldbuilding} (target distances $\{d_{ij}\}$ and the configuration in rings) would easily allow to get target positions $p_i^*$ for each agent and then drive the agents to such targets. However, the main drawbacks of this approach include: 1) The requirement of a global coordinate system for the agents; 2) interactions among the agents are sometimes desirable to enhance control performance or address additional objectives such as formation shape-keeping \citep{Oh2015}; 3) the agents will follow, in general, a shorter path in the distance-based approach, especially if the surface has any symmetry since the final positions will be those that, satisfying the constraints, are closer to the initial conditions; 4) in terms of failures or loss of agents, the system can be better reconfigured when there exists a topology between nodes and the formation is defined in terms of distances. Therefore, we can say that it offers a more robust behavior. 
\end{rmk11}

Let us define the error functions as
\begin{equation} \label{eq:errorij}
    e_{ij}=d_{ij}^2-d_{ij}^{*2},
\end{equation}
i.e., the error between the target distances and the square norm for the edge $ij$, or equivalently $e_k=\|z_k\|^2-d_k^{*2}$. 
Let us also define the following stack vectors $p^\top=(p_1^\top,\dots,p_N^\top)$, $f_{\mathcal{S}}^\top(p)=(f_{\mathcal{S}}(p_1),\dots,f_{\mathcal{S}}(p_N))$, and $e^\top=(\dots,e_k,\dots)$. Then, \eqref{eq:potential} can be rewritten as 
\begin{equation}
W=W_1+W_2,
\end{equation}
where 
\begin{align}
  W_1&=\frac{\kappa_1}{4} e^\top e \\
  W_2&=\frac{\kappa_2}{4}f_{\mathcal{S}}^\top(p)f_{\mathcal{S}}(p). 
\end{align}
With the above definitions, the overall system dynamics can be rewritten as
\begin{equation}\label{eq:dynamicsAll}
    \dot{p}=-\kappa_1R^\top(z) e-\kappa_2J_\mathcal{S}^\top(p) f_{\mathcal{S}}(p),
\end{equation}
where $R(z)\equiv J_{f_\mathcal{G}}(p)$ is the rigidity matrix of the graph $\mathcal{G}$ and $J_\mathcal{S}$ is the Jacobian matrix of the function $f_{\mathcal{S}}(p)$. Note that $R(z)$ has a row for each edge and 3 (in $\mathbb{R}^3$) columns for each vertex, so that the $k$-th row of $R(z)$ corresponding to the $k$-th edge of $\mathcal{E}$ connecting vertices $i$ and $j$ is
$$[0 \dots 0 \ (p_i-p_j)^\top 0 \dots 0 \ (p_j-p_i)^\top 0 \dots 0].$$
$J_\mathcal{S}$ has a block diagonal structure, such that each diagonal block $i$, $i=1,\dots,N$, is $p_i^\top Q_1$.

\subsection{Stability analysis}

In this section, we analyze the equilibria and stability of the system \eqref{eq:sysDynamics} under the control law \eqref{eq:controlLaw}. 
Some manipulations will be useful in the following analysis. The product $R^\top(z) e$ can be rewritten as $(\bar{E}(p)\otimes I_3)p$ \citep{Anderson2014}, where  $\bar{E}(p)=H^\top E(p) H$, being $E(p)$ a  diagonal matrix defined as $E(p)=diag(\dots,e_k,\dots)$. Similarly, we can define $F_\mathcal{S}(p)=diag(\dots,f_\mathcal{S}(p_i),\dots)$ so that the product $J_\mathcal{S}^\top(p) f_{\mathcal{S}}(p)$ can be rewritten as 
$(F_\mathcal{S}(p)\otimes Q_1)p$. Then, \eqref{eq:dynamicsAll} is equivalent to
\begin{equation}\label{eq:dynamicsAll2}
    \dot{p}=-\kappa_1(\bar{E}(p)\otimes I_3)p-\kappa_2(F_\mathcal{S}(p)\otimes Q_1)p,
\end{equation}

The following analysis will study the stability of the multi-agent system \eqref{eq:sysDynamics} under the control law \eqref{eq:controlLaw}.

\begin{prop3} \label{lemma1}
The multi-agent system \eqref{eq:sysDynamics} with control law \eqref{eq:controlLaw} has an equilibrium set $\mathcal{M}_d$ defined by
\begin{equation} \label{eq:setDesired}
    \mathcal{M}_d=\{e=0, f_\mathcal{S}(p)=0\}
\end{equation}
corresponding to the control objective, i.e., acquisition of the desired formation defined by the prescribed distances $d_{ij}^*$ and placement of the agents over the virtual surface $\mathcal{S}$ defined by \eqref{eq:shieldSurface}.
\end{prop3}
\begin{proof}
The control objective is satisfied if and only if
\begin{enumerate}
    \item $d_{ij}=d_{ij}^*, \ \forall (i,j)\in\mathcal{E}$
    \item $f_\mathcal{S}(p_i)=0, \ \forall i=1,\dots,N.$
\end{enumerate}
Then, the Lyapunov function \eqref{eq:potential} is 0 if and only if the control objective is achieved. Moreover, the time derivative of the Lyapunov function along the system solution is
$$\dot{W}=\big(\frac{\partial W_1}{\partial p}+\frac{\partial W_2}{\partial p}\big)\dot{p}=\big(\frac{\kappa_1}{2}e^\top\frac{\partial e}{\partial p}+\frac{\kappa_2}{2}f_{\mathcal{S}}^\top(p)\frac{\partial f_{\mathcal{S}}(p)}{\partial p}\big)\cdot \dot{p}.$$
Note that by definition the partial derivatives are the Jacobian matrices defined above, i.e., $R(z)$ and $J_\mathcal{S}$, respectively, then it holds that
$$\dot{W}=-\big(\kappa_1 e^\top R(z)+\kappa_2 f_{\mathcal{S}}^\top(p) J_{\mathcal{S}}(p)\big)\big(\kappa_1R^\top(z) e+\kappa_2J_\mathcal{S}^\top(p) f_{\mathcal{S}}(p)\big),$$
hence 
$$\dot{W}=-\|\kappa_1R^\top(z) e+\kappa_2J_\mathcal{S}^\top(p) f_{\mathcal{S}}(p)\|^2\leq 0.$$
Then, the Lyapunov function \eqref{eq:potential} is not increasing along the system solutions, and $\dot{W}=0$ at the equilibrium set defined in \eqref{eq:setDesired}. 
Hence, the proof is completed.
\end{proof}

\begin{rmk2}
Note that the complete set of equilibria of \eqref{eq:dynamicsAll} is defined by
\begin{equation} 
    \mathcal{M}=\{p: \ \kappa_1R^\top(z) e+\kappa_2J_\mathcal{S}^\top(p) f_{\mathcal{S}}(p)=0\},
\end{equation}
such that $\mathcal{M}_d\subset\mathcal{M}$.

The fact that other equilibria sets exist also occurs in the problem of rigid formations  \citep{Krick2009,Sun2015}, and the conditions to facilitate the demonstration of the local stability relies on imposing the conditions of minimal and infinitesimal rigidity of the framework (see Section \ref{sec:graphRigidity}). However, this applies when the formation is realized in the state space $\mathbb{R}^2$/$\mathbb{R}^3$. In the setup presented in this paper, two main differences makes that the results are not applicable. First, the state of the agents $p_i\in\mathbb{R}^3$ but the formation is embedded in a virtual surface $\mathcal{S}$ of dimension 2. And secondly, constraining the formation to $\mathcal{S}$ makes that the concept of infinitesimally rigid cannot be applied as such. Actually, the rigid body motions corresponding to the translation along the axes can no longer occur, and the rotations about one or more axes depend on the symmetries of $\mathcal{S}$.

Furthermore, note that an augmented matrix and state vector can be constructed as:
\begin{align}
    J_{R\mathcal{S}}^\top(p,z)&=\begin{pmatrix}\kappa_1R^\top(z) & \kappa_2J_{\mathcal{S}}^\top(p)\end{pmatrix} \label{eq:Jrs}\\
    \xi^\top(p,z)&=\begin{pmatrix} e(z)^\top & f_\mathcal{S}^\top(p)\end{pmatrix},
\end{align}
such that studying the rank of $J_{R\mathcal{S}}(p,z)\in \mathbb{R}^{(N_e+N)\times 3N}$ in this setup is equivalent to study the rank of the rigidity matrix in the classical problem of rigid formations. Actually, the non-zero elements of $J_{\mathcal{S}}$ can be seen as a square distance from the $N$ nodes to a \textit{virtual node} at the origin weighted by the matrix $Q_1$, and then the number of edges (real plus \textit{virtual}) is $N_e+N$. According to Proposition \ref{prop:prop1}, $N_e+N$ belongs to the interval $[3N-2,4N-6]$. Note that if we include this virtual node (labeled as $0$ and corresponding to the origin) in the counting of vertices, $\mathcal{V}'=\mathcal{V}\cup {0}$, such that the number of nodes is $N'=N+1$ and then the number of edges is $N'_e\in[3N'-5,4N'-9]$. Then, the framework is not minimally rigid under this transformation of the problem, and this can also be inferred from the results of Proposition \ref{prop:prop1}, as we will discuss next in the paper.
\end{rmk2}

We next analyze the rank of $J_{R\mathcal{S}}(p,z)$. Note that since the quadric surface \eqref{eq:shieldSurface} is assumed to be expressed in normal form,  $Q_1$ has diagonal form such that $Q_1=diag(q_1,q_2,q_3)$ (see Table \ref{tab:shieldexample}).
\begin{prop5} \label{lem2}
The rank of $J_{R\mathcal{S}}(p,z)$ in \eqref{eq:Jrs} for a rigid framework defined by a Delaunay Triangulation and embedded in a surface $\mathcal{S}$ defined as \eqref{eq:shieldSurface} is at least $3N-3$. Moreover, when then number of edges of $\mathcal{G}$, is such that $N_e\geq 2N$, then it holds that
\begin{equation}
   rank(J_{R\mathcal{S}})=3N-s, 
\end{equation}
where $s$ reflects the symmetries of $\mathcal{S}$ such that
\begin{equation} \label{eq:casess}
    s=\begin{cases}
    0 &\text{ if } q_i\neq q_j, \ \forall i\neq j, \ i,j\in\{1,2,3\} \\
    1 & \text{ if } \exists i,j,k\in\{1,2,3\} \ q_i=q_j,q_i\neq q_k \ i\neq j \neq k, \\
    3 & \text{ if } q_i=q_j \ \forall i,j\in\{1,2,3\}.
\end{cases}
\end{equation}
\end{prop5}
\begin{proof}
The proof can be found in the Appendix.   
\end{proof}

We finally present the main result of this section regarding stability based on the previous developments.

\begin{thm2} \label{thm:thm2}
The multi-agent system \eqref{eq:sysDynamics} for a given shield model described by \eqref{eq:shieldSurface}, the graph constructed such that the target formation is a Delaunay triangulation over $\mathcal{S}$, and the control law \eqref{eq:controlLaw}, is locally asymptotically stable at the desired relative positions $d_{ij}^*$ over the surface $\mathcal{S}$ corresponding to $e=0$ and $f_{\mathcal{S}}(p)=0$.
\end{thm2}
\begin{proof}
The proof can be found in the Appendix.
\end{proof}

\subsection{Truncated surfaces} \label{sec:truncatedSurf}

To deal with the constraints on the $z$ axis, which may result in truncated surfaces, we introduce an additional term in the control law by adapting classical techniques for obstacle avoidance \citep{Khatib1986}. More specifically, a repulsive potential field is defined to avoid that agents' trajectories cross the plane $z=0$: 
\begin{equation} \label{eq:repU}
    U_r(p_z)=\begin{cases}
    \frac{\kappa_3}{2}\Big(\frac{1}{p_z}-\frac{1}{\epsilon}\Big)^2 & \text{ if } p_z\leq \epsilon \\
    0 & \text{ if } p_z>\epsilon,
\end{cases}
\end{equation}
where $\epsilon>0$ acts as a threshold to activate the repulsive potential field, and usually takes small values. The corresponding control term is
\begin{equation} \label{eq:u3}
    u_{r_i}=-\nabla U_r(p_{z,i})=\begin{cases}
    \kappa_3\Big(\frac{1}{p_{z,i}}-\frac{1}{\epsilon}\Big)\frac{1}{p_{z,i}^2} & \text{ if } p_{z,i}\leq \epsilon \\
    0 & \text{ if } p_{z,i}>\epsilon.
\end{cases}
\end{equation}
Assuming that the initial conditions are such that $p_{z,i}(0)\geq 0$, \eqref{eq:u3} guarantees that $p_{z,i}(t)>0, \ \forall t$. Then, the control law \eqref{eq:controlLaw} is transformed into
\begin{align} \label{eq:controlLaw2}
    u_i=&-\kappa_1\sum_{j\in\mathcal{N}_i}(d_{ij}^2-d_{ij}^{*2})(p_i-p_j)-\frac{\kappa_2}{2}f_{\mathcal{S}}(p_i)\frac{\partial f_{\mathcal{S}}(p_i)}{\partial p_i} \nonumber \\
    &+u_{r,i}.
\end{align}

\begin{rmk4}
    If the surface $\mathcal{S}$ has another constraint on $z$ such as some of the examples presented in Table \ref{tab:shieldexample} ($p_z\leq h$), the problem is solved adding a new term similar to \eqref{eq:u3} but replacing $p_{z,i}$ by $h-p_{z,i}$ and assuming that $p_{z,i}(0)\leq h$.
\end{rmk4}

\begin{rmk10}
The introduction of the repulsive potential field of the form \eqref{eq:repU} has been used for decades and represents a simple solution to avoid collisions with obstacles and with other agents. However, repulsive potential fields have the drawback that can generate additional local minima in which the agent can be trapped. To avoid this, there exist several solutions, including the introduction of a uniformly bounded perturbation term, tangent to the level curves of the repulsive potential \eqref{eq:repU}, when the trap situation occurs \citep{Qu2009}. Also, the definitions of repulsive potential functions as control barrier functions, which serve as a method for providing safety guarantees and provide more elegant solutions, have been addressed recently \citep{Singletary2021}. \\
However, in the case of \eqref{eq:repU} in the framework presented in this paper, this corresponds with some $\hat{p}$ such that 
$$\frac{\partial(W_1+W_2)(\hat{p})}{\partial \hat{p}}=-\frac{\partial U_{r} (\hat{p})}{\partial \hat{p}},$$
where $\hat{p}$ is such that $0<\hat{p}_{z,i}<\epsilon$ for some $i$ at the lower level of the shield, and it constitutes a small perturbation of the desired formation since the region where $U_r$ acts (defined by $\epsilon$) is small compared to the shield dimensions. Hence, the aforementioned solutions have not been considered so far.    
\end{rmk10}

\section{Simulation and experimental results} \label{sec:simulations}
\subsection{Simulation example 1}
Let us consider a team of $N=50$ agents and a semi-ellipsoid as desired shield shape as follows:
\begin{equation} \label{eq:ellipsoid}
    \frac{x^2}{10^2}+\frac{y^2}{15^2}+\frac{z^2}{12^2}=1.
\end{equation}
The execution of Algorithm \ref{alg:shieldbuilding} gives, as a result, a value for the distance between neighbors of $d=5.154$, and a distribution of levels as shown in Table \ref{tab:ellipsoide}. The lowest level denoted as $h=0$ is a practical simplification since there exists the repulsive potential field at $z=0$ (see Section \ref{sec:truncatedSurf}), and then the height of this level should be $h>\epsilon$. Since $\epsilon$ is a small value, the effect over the results is not significant.
\begin{table}[th!]  
\renewcommand{\arraystretch}{1.2}
\centering
\begin{tabular}{||c | c c c c c||} 
 \hline  
 $h_k$ & $0^*$ & 5.078 & 8.422 & 10.750 & 11.938 \\ [0.5ex] 
 \hline 
 $n_k$ &	16  & 14 & 11 & 7 & 2\\ [0.5ex]
 \hline 
 $d_k$ &	4.958  & 5.134 &  5.137 & 5.042 & 4.039\\ [0.5ex]
 \hline
\end{tabular}
\caption{Values obtained for $h_k$, $N_k$, and $d_k$ in Algorithm \ref{alg:shieldbuilding} for the semi-ellipsoid defined in \eqref{eq:ellipsoid}  and $N=50$.}
\label{tab:ellipsoide} 
\end{table}
\begin{figure}[th]
  \centering
  \parbox{\figrasterwd}{
    \parbox{.55\figrasterwd}{%
      {\includegraphics[width=\hsize]{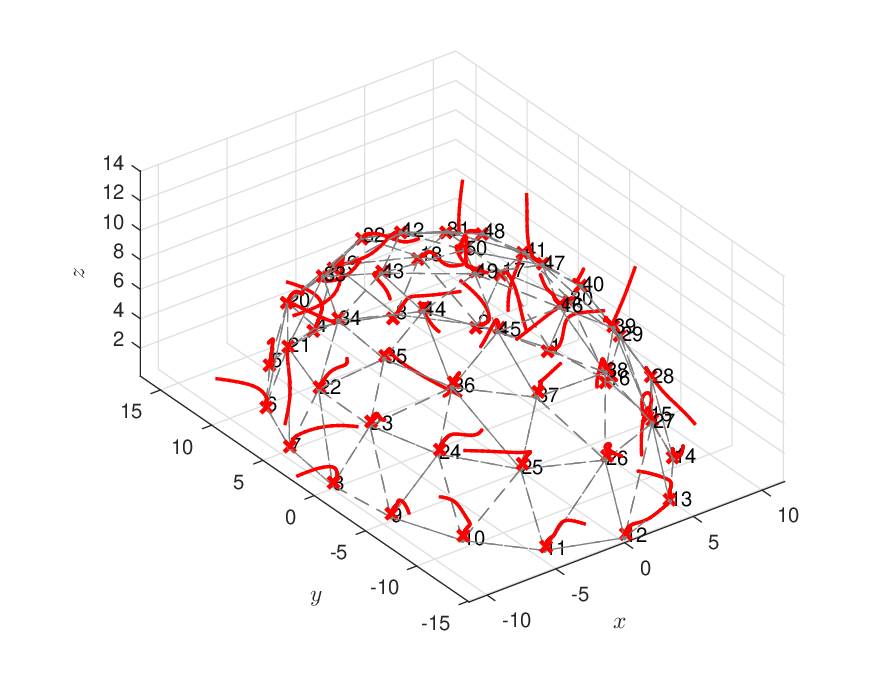}}
    }
    \hskip1em
    \parbox{.45\figrasterwd}{%
      {\includegraphics[width=\hsize]{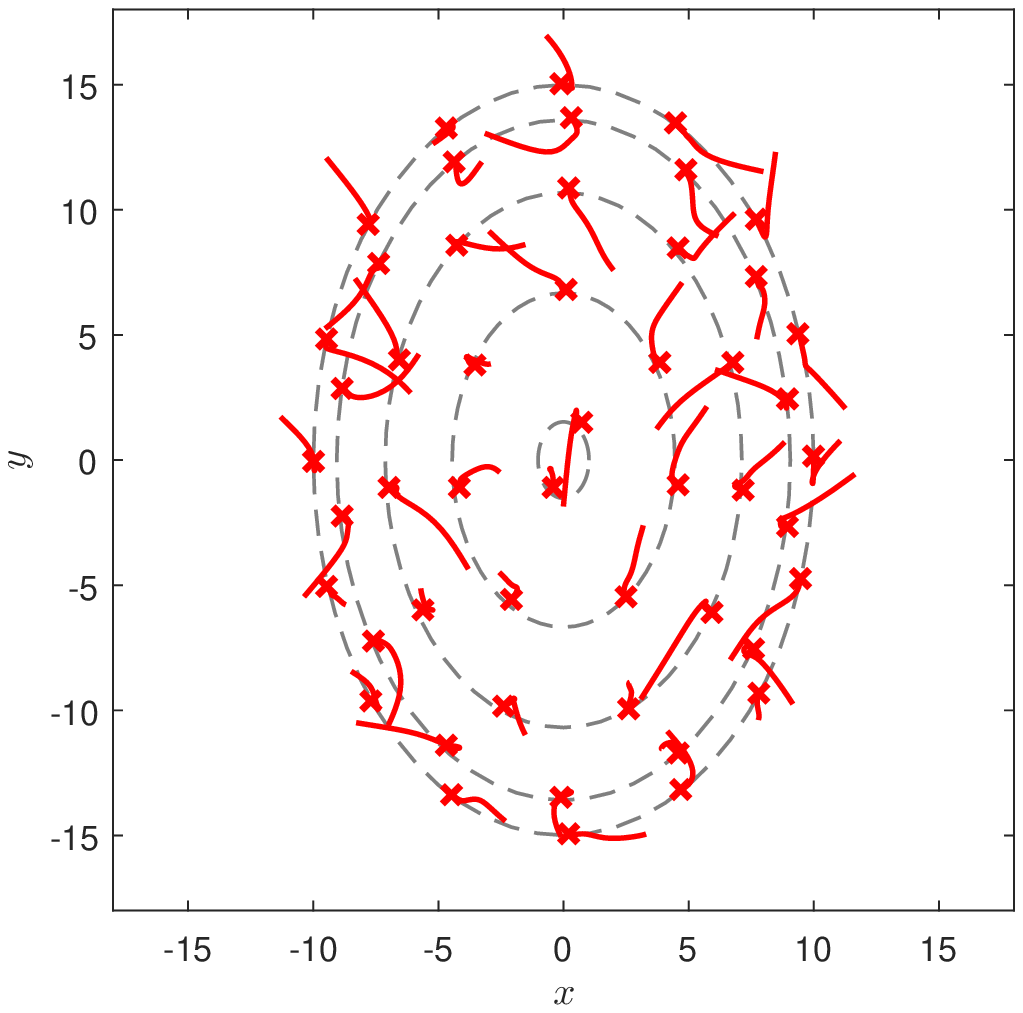}}
    }
    }
  \caption{3D view (left) and projection over the XY plane (right) of the trajectories of the system (in red) for Example 1. Red crosses represent positions at which errors $e_{ij}$ are 0.}
  \label{fig:ellipsoid}
\end{figure}

The left hand side of Figure \ref{fig:ellipsoid} shows the trajectories of the agents in the 3D space when the initial conditions are generated randomly but with a bound such that $|d_{ij}(0)-d_{ij}^*|\leq 7$ for all neighboring agents $i$ and $j$, and $|f_\mathcal{S}(p_i(0))|\leq 7\|Q_1\|=0.07$. The control law \eqref{eq:controlLaw2} with feedback gains $\kappa_1=0.1, \ \kappa_2=10^3, \ \kappa_3=10^{-3}$ is applied. The topology of the system in the form of Delaunay triangulation is also depicted. The right hand side of Figure \ref{fig:ellipsoid} shows the projection of those trajectories over the XY plane. Dashed elipses Note that the agents converge to the surface and they acquire the desired target distance between neighboring nodes.
\begin{figure}[th!]
    \centering
    \parbox{\figrasterwd}{
    \parbox{.6\figrasterwd}{%
      {\includegraphics[width=\hsize]{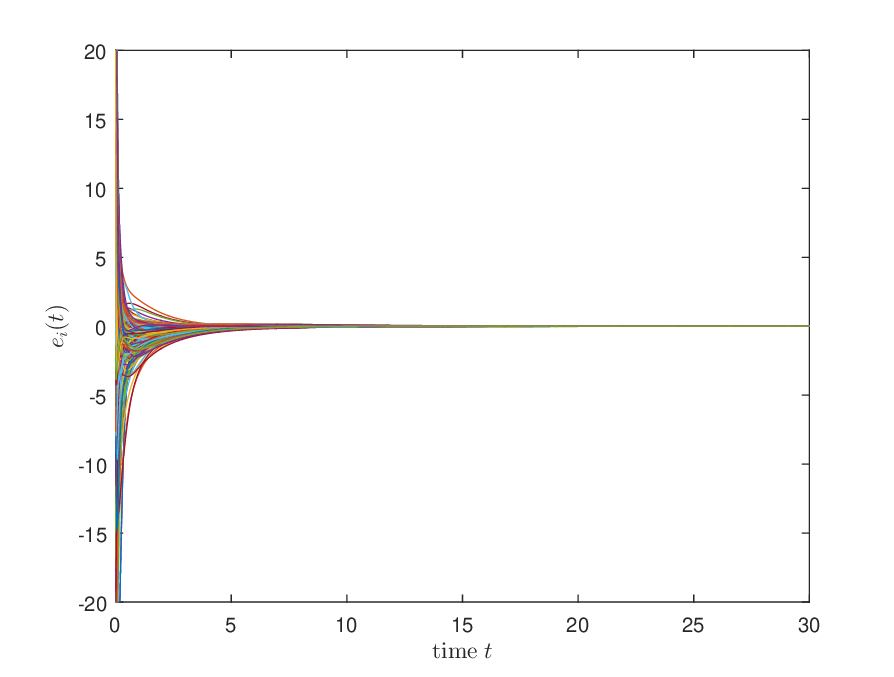}}
    }
    \hskip1em
    \parbox{.35\figrasterwd}{%
      {\includegraphics[width=\hsize]{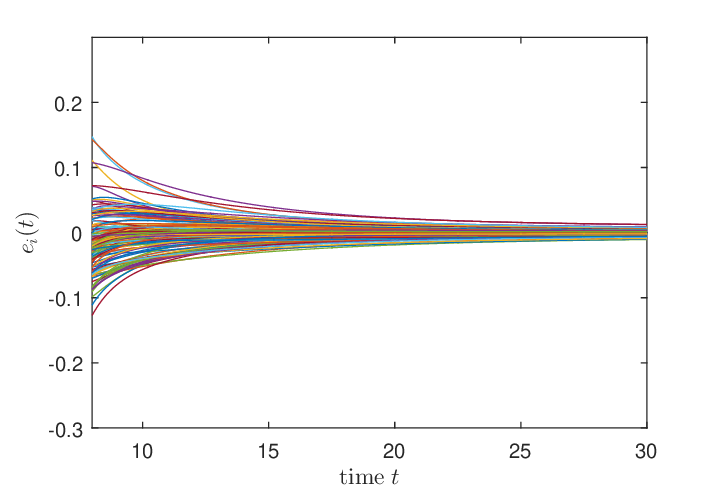}}
    }
    }
    \caption{Evolution of the error to target distances over time for Example 1. The right-hand side shows a zoom for the interval of time $t\in[8,30]$.}\label{fig:evst}
\end{figure}

\begin{figure}[th!]
    \centering
    \includegraphics[width=0.9\hsize]{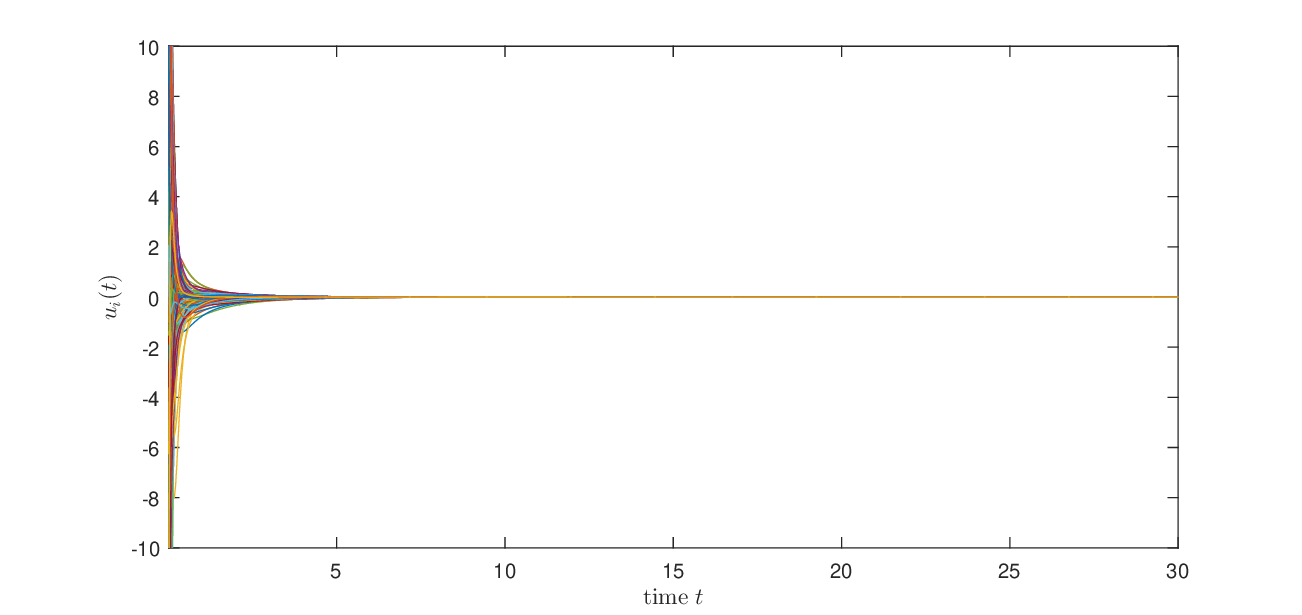}
    \caption{Control signals $u_i(t)$ according to \eqref{eq:controlLaw2}.}\label{fig:u}
\end{figure}

Figure \ref{fig:evst} shows the evolution of the error $e(z)$ over time, where a zoom for the interval of time $t\in[8,30]$ is depicted on the right-hand side. Note that the error for edges converges to 0. 
Figure \ref{fig:u} shows the control signals computed as in \eqref{eq:controlLaw2}. Note that they also converge to 0 asymptotically. 

\begin{figure}[th!]
    \centering
    \includegraphics[width=1\hsize]{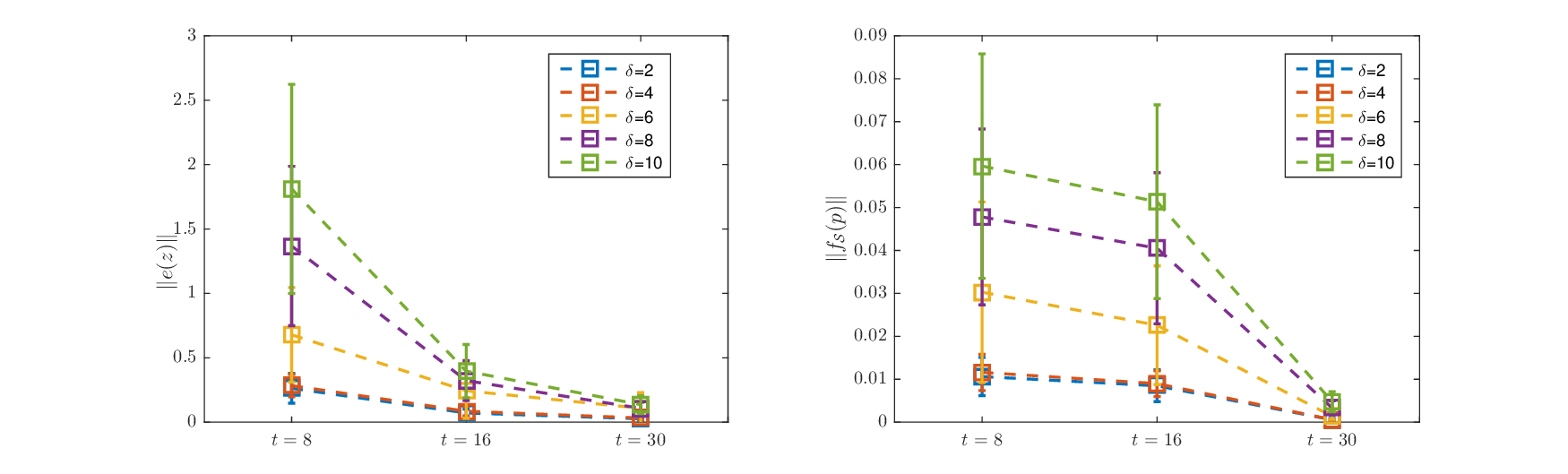}
    \caption{Mean and standard deviation at $t=8, 16, 30$ for different values of $\delta$. Left: Norm of the overall system error $e(z)$. Right: Norm of the overall system function $f_\mathcal{S}(p)$.}\label{fig:statistics}
\end{figure}

Finally, a statistical study has been performed to analyze the influence of the initial conditions over the performance of the system. More specifically, the norm of errors for the whole system $e(z)$ and $f_\mathcal{S}(p)$ has been computed for a set of experiments with initial conditions such as $|d_{ij}(0)-d_{ij}^*|\leq \delta$ and $|f_\mathcal{S}(p_i(0))|\leq \|Q_1\|\cdot \delta=0.01\delta$, with $\delta=2, 4, 6, 8, 10, 14$. For each $\delta$, 5 simulations with random initial conditions have been performed. The feedback gains and duration of experiments are the same as described above. The mean and standard deviation (SD) at $t=0$ are shown in Table \ref{tab:statistics}. In Figure \ref{fig:statistics} the graphs show the mean and standard deviation (with bars) for $\|e(z)\|$ and $\|f_\mathcal{S}(p)\|$, respectively, at $t=8, 16, 30$ for $\delta\in\{2,4,6,8,10\}$. In all cases, the reduction of the errors $\|e(z)\|$ and $\|f_\mathcal{S}(p)\|$ at $t=8$ are over the $99.7 \%$ and $98.3 \%$, respectively. For $\delta=14$ the performance is not acceptable, specially for $\|e(z)\|$, since at $t=8,16,30$ mean values of 28.57, 10.05, and 3.57, respectively, are obtained, around 20 times greater than for $\delta=10$.
\begin{table}[th!]  
\renewcommand{\arraystretch}{1.2}
\centering
\begin{tabular}{|| c | c | c c c c c c||} 
 \hline  
 & $\delta$ & $2$ & 4 & 6 & 6 & 10 & 14 \\ [0.5ex] 
 \hline 
 \multirow{2}{5em}{$\|e(z(0))\|$} &Mean &	94.3  & 205.8 & 372.8 & 510.5 & 817.6 & 1259.8\\ [0.5ex]
  & SD &	10.65  & 12.69 &  36.42 & 43.31 & 109.92 & 79.33\\ [0.5ex]
 \hline
 \multirow{2}{4em}{$\|f_\mathcal{S}(p(0))\|$} & Mean &	0.712  & 1.446 & 2.291 & 2.942 & 3.865 & 5.331\\ [0.5ex]
  & SD &	0.039  & 0.093 &  0.153 & 0.162 & 0.559 & 0.955\\ [0.5ex]
 \hline
\end{tabular}
\caption{Initial values for the mean and standard deviation for $\|e(z)\|$ and $\|f_\mathcal{S}(p)\|$.}
\label{tab:statistics} 
\end{table}

\subsection{Simulation example 2}
\begin{figure}[ht!]
  \centering
  \parbox{\figrasterwd}{
    \parbox{.5\figrasterwd}{%
      {\includegraphics[width=\hsize]{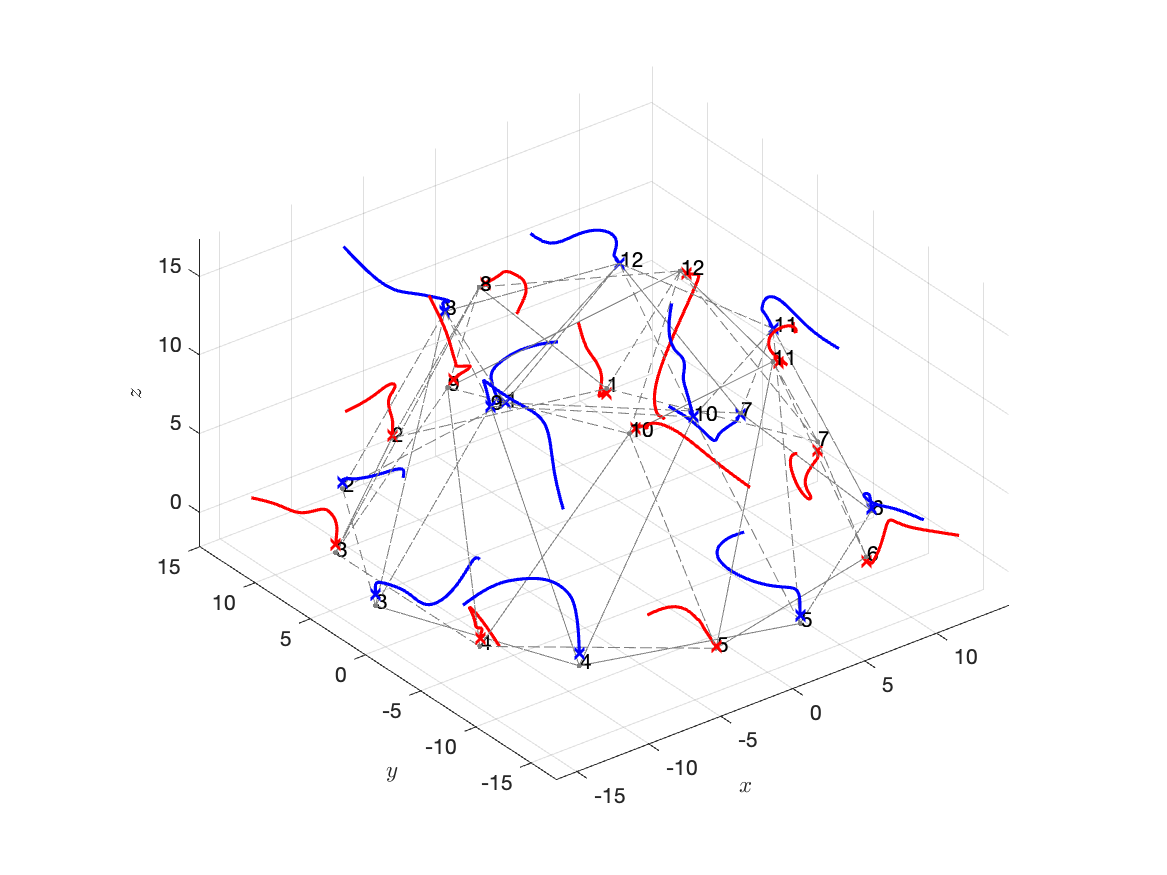}}
    }
    \hskip1em
    \parbox{.5\figrasterwd}{%
      {\includegraphics[width=\hsize]{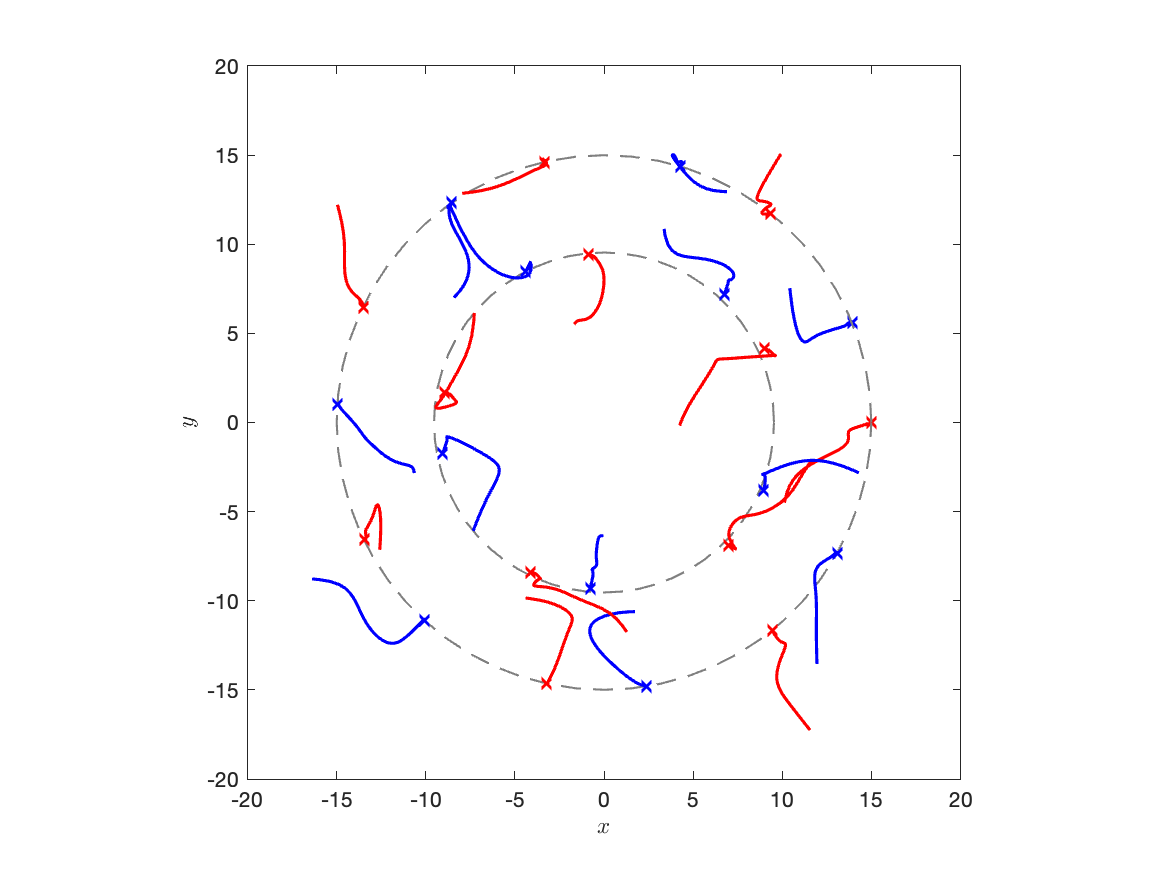}}
    }
    }
  \caption{3D view (left) and projection over the XY plane (right) of the trajectories of the system (in red) for two different initial conditions (Example 2). Dashed circles represent the rings of height $h=0$ and $h=3.421$.}
  \label{fig:sphere}
\end{figure}
To illustrate the results when the surface presents symmetries, let us consider the case of a semi-sphere with $R=15$ and a multi-agent system with $N=12$. In this case, Algorithm \ref{alg:shieldbuilding} distributes the agents in two rings with 7 and 5 drones at heights $h=0^*$ (similar comments as the previous example applies) and $h=3.421$, respectively. The left of Figure \ref{fig:sphere} shows the trajectories and the topology of the system in the 3D space for two different initial conditions. For the data in red, at $t=0$, the norm of the relative errors' vector is $\|e(z(0))\|=684.68$ and  $\|f_\mathcal{S}(p(0))\|=1.861$. At $t=15$, these values are reduced to 0.003 and $5,8\times 10^{-4}$, respectively.  
For the data in blue, $\|e(z(0))\|=665.50$ and $\|f_\mathcal{S}(p(0))\|=1.706$, and at $t=15$, $\|e(z(15))\|=0.0021$ and $\|f_\mathcal{S}(15)\|=1.8\times 10^{-4}$. Then, the control objective is achieved in both cases but the final positions differ (there exists a rotation) influenced by the initial conditions. The projection of the trajectories over the XY plane is depicted on the right of Figure \ref{fig:sphere}. Dashed circles represent the rings of height $h=0$ and $h=3.421$ that Algorithm \ref{alg:shieldbuilding} computes to ensure an \textit{almost uniform} distribution of the nodes.

\subsection{Real-time experiment}

\begin{figure}[th]
   \centering
      {\includegraphics[width=\hsize]{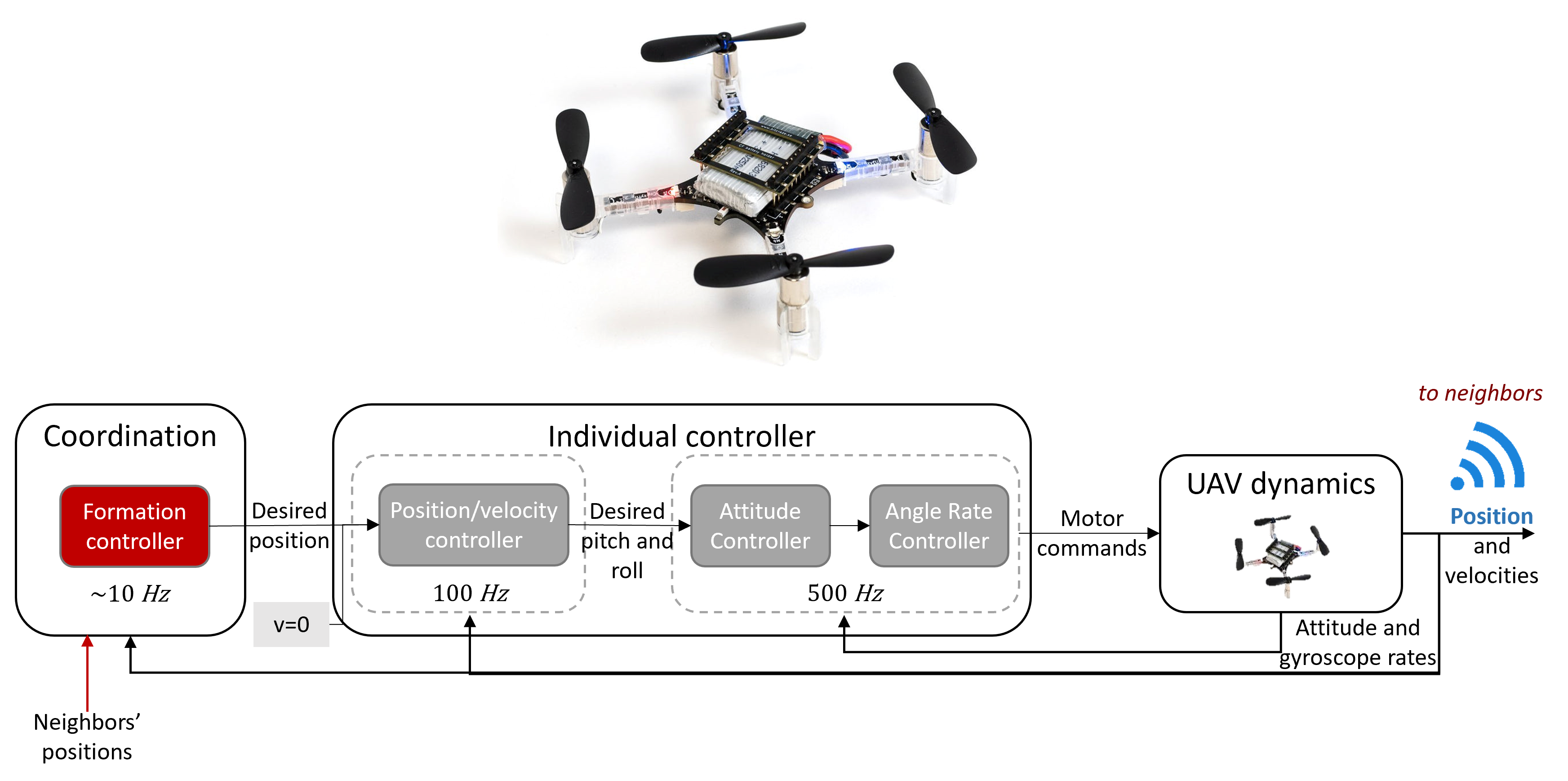}}
  \caption{Crazyflie 2.1 (top), hierarchical control architecture (bottom).}
  \label{crazyflie}
\end{figure}

The proposed strategy has also been tested over the experimental platform described in \citet{Manas2023}, which supports different autonomous robots including UAVs. Demo videos of the platform can be found at \url{https://www.youtube.com/@roboticpark4354}. A team of 12 micro-aerial quadcopter Crazyflie 2.1 \citep{Giernacki2017} (see Figure \ref{crazyflie} top) has been used for the experiment presented in this paper, 6 of which are physical robots and 6 are virtual robots.  The agents interact with each other as they all were real thanks to the platform developed in ROS 2. 

The physical robot has a STM32F405 microcontroller and a Bluetooth module that allows the communication. The Crazyflie uses its own positioning system, the Lighthouse \citep{Taffanel2021}, which is based on infrared laser and enables the Crazyflie to  calculate its own position onboard with a precision of 1 mm. The control architecture follows a hierarchical scheme (see Figure \ref{crazyflie} bottom). The dynamics of the UAV can be classified into the trajectory dynamics and the attitude and the angle dynamics \citep{Bayezit2012} Therefore, the individual control architecture can follow a cascade structure \citep{Dong2016}, where the inner loop stabilizes the attitude and runs at a higher frequency (500 Hz), and the outer loop controls the position and velocity of the drone running at 100 Hz. The proposed controller in this paper represents another level of the control scheme (\textit{coordination}), which provides a goal position to the individual controller and runs at a frequency of 10 Hz. This hierarchical structure with different sampling frequencies allows us to consider an approximate model of the quadrotor UAV dynamics in the outer level as in \eqref{eq:sysDynamics}.
\begin{figure}
    \centering
    \includegraphics[width=0.8\hsize]{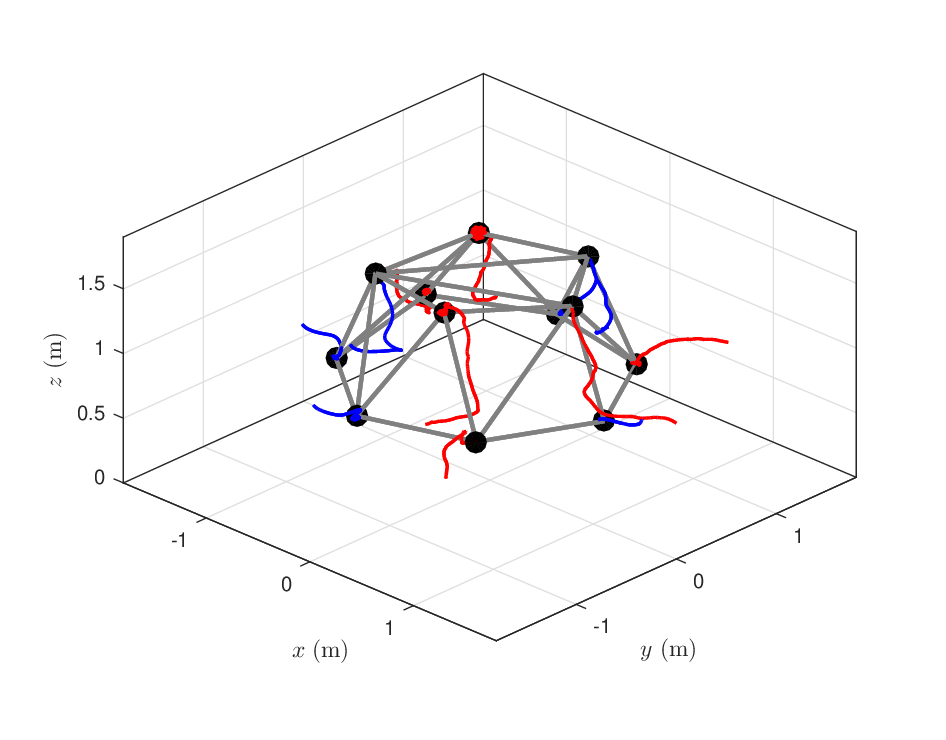}
    \caption{3D view of the trajectories of the team of 12 Crazyflies 2.1: physical robots (red), virtual robots (blue).}
    \label{fig:trajectoriesReal}
\end{figure}
\begin{figure}[ht!]
  \centering
  \parbox{\figrasterwd}{
    \parbox{.5\figrasterwd}{%
      {\includegraphics[width=\hsize]{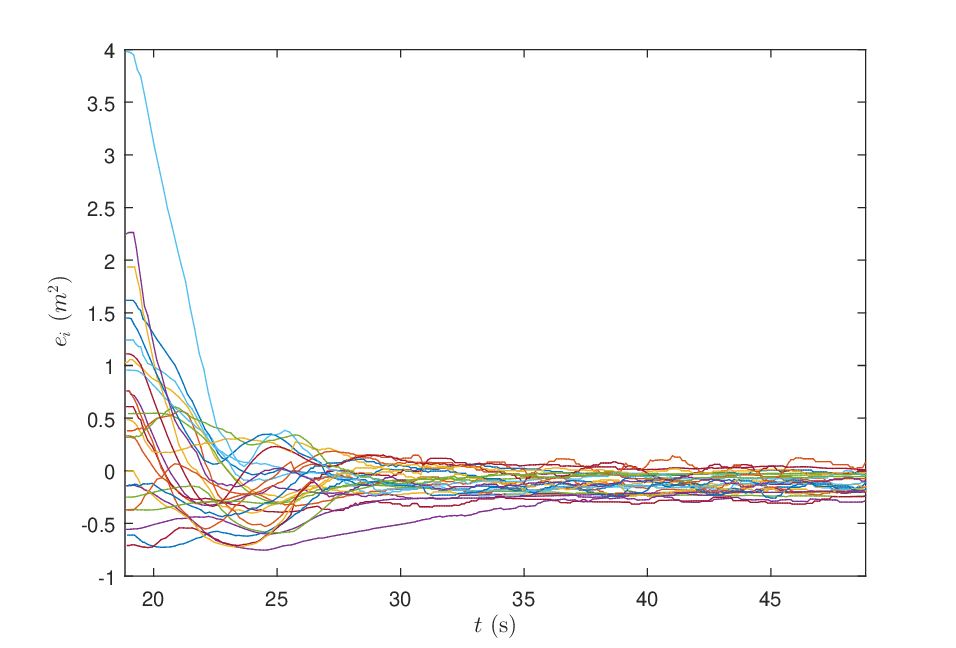}}
    }
    \hskip1em
    \parbox{.5\figrasterwd}{%
      {\includegraphics[width=\hsize]{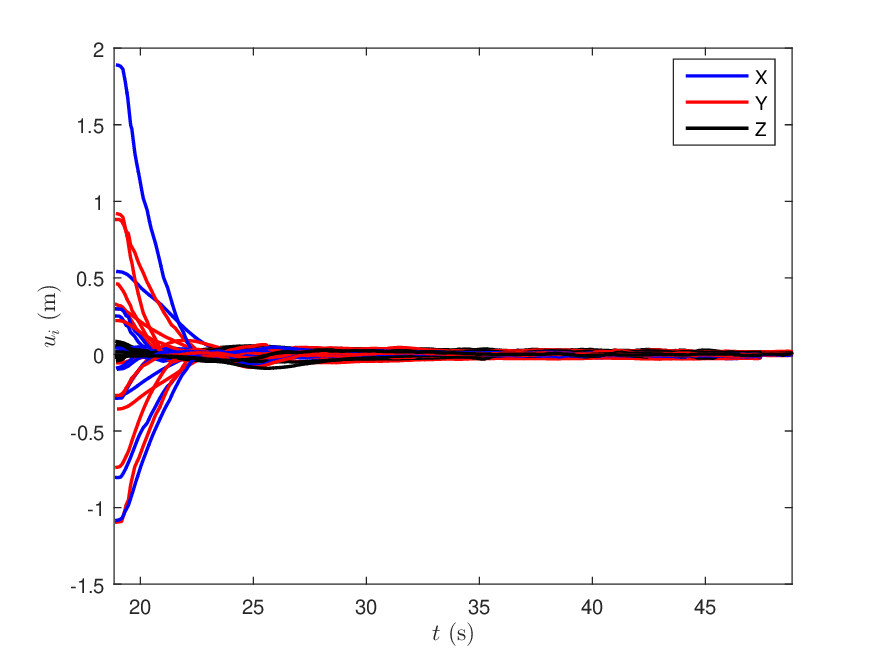}}
    }
    }
  \caption{Evolution of the error to target distances for the 26 edges of the graph (left), and control signals of the coordination controller (right).}
  \label{fig:experResults}
\end{figure}

The virtual surface in this experience is a semisphere with $R=1$ m, whose center is at $(0,0,0.8)$ m. The execution of Algorithm \ref{alg:shieldbuilding} provides a value of $d=0.97$ m, and a distribution of agents in two rings with $n_b=7$ and $n_1=5$ at heights $h_0=0.8$ m and $h=1.57$ m with parameters $d_0=0.9$ m and $d_1=0.80$ m, respectively. The number of edges of the resulting triangulation is 26. The robots are first commanded to move to the plane $z_0=0.8$ m, and then the coordination controller starts working. The trajectories of the robots are depicted in Figure \ref{fig:trajectoriesReal}, where red lines represent the physical robots and blue lines the virtual drones. The left of Figure \ref{fig:experResults} shows the evolution of the error of the formation over time, where it is clear that the multi-agent system converges to the desired formation, and the signals of the coordination controller an depicted on the right. Different colors are used for $u_{x,i}, u_{y,i}, u_{i,z}$: blue, red, and black, respectively. A video of this experiment can be found at: \url{https://youtu.be/j8SpkPp_5zs}.

\section{Conclusions} \label{sec:conclusions}
In this paper, we have studied the formation shape control problem of a set of agents moving in the 3D space. They should achieve a formation in such a way that they form a shield and are distributed over a virtual surface modeled as a quadric in normal form. The potential application is the protection of an area of interest and the monitoring of external threats. An algorithm has been proposed to guarantee an \textit{almost uniform} distribution of the nodes and the network configuration in the form of a Delaunay triangulation. A method to test if each triangle is Delaunay has been designed, so that it can be executed locally. Moreover, a distributed control law has been proposed to guarantee the achievement of the control objective. Although the conditions of minimal and infinitesimal rigidity of the framework do not hold in our setting, we have been able to provide proofs of local stability. The simulation and experimental results have shown that the proposed control method yields asymptotic stability of the desired formation.

Although the analysis of this paper was centered on single integrator agents, we have been able to apply it to UAVs thanks to a hierarchical control architecture. However, the extension of the proposed approach to more detailed UAV models will be part of future work. Also, we will study switching topologies and the design of strategies to handle disturbances and failures in the system (loss of agents or sensing capacities).

\section*{Acknowledgments}
This work was supported in part by Agencia Estatal de Investigaci\'{o}n (AEI) under the Project PID2020-112658RB-I00/AEI/10.13039/501100011033.

\section*{Appendix. Proofs}

\subsection*{Proof of Lemma \ref{lem:lem1}}
The equation \eqref{eq:CC1} can be rewritten as 
$$(I-P)\Bigg(\begin{pmatrix}\|p_A\|^2 \\ \|p_B\|^2 \\ \|p_C\|^2\end{pmatrix}-2\begin{pmatrix}p_A^\top \\ p_B^\top \\ p_C^\top\end{pmatrix}m_{ABC}\Bigg)=\mathbf{0},$$
where $P$ is a permutation matrix given by
$$P=\begin{pmatrix}0 & 1 & 0 \\ 0 & 0 & 1 \\ 1 & 0 & 0\end{pmatrix}.$$
Then, $I-P$ is a Laplacian matrix, and hence, $(I-P)\mathbf{1}=\mathbf{0}$, and it follows that
\begin{equation} \label{eq:eqm1}
    \begin{pmatrix}\|p_A\|^2 \\ \|p_B\|^2 \\ \|p_C\|^2\end{pmatrix}=2\begin{pmatrix}p_A^\top \\ p_B^\top \\ p_C^\top\end{pmatrix}m_{ABC}+2\gamma \begin{pmatrix}1 \\ 1 \\ 1\end{pmatrix}
\end{equation}
for any $\gamma$. Additionally, since $m_{ABC}=(p_{x,ABC},p_{y,ABC},p_{z,ABC})\in f_\pi$, it holds that
$$a\cdot p_{x,ABC}+b\cdot p_{y,ABC}+c\cdot p_{z,ABC}+d=0.$$
The parameters of $f_\pi$ in \eqref{eq:planeDef} are defined in \eqref{eq:ab_plane}-\eqref{eq:cd_plane}. Furthermore, the normal vector of $f_\pi$ in \eqref{eq:planeDef} is $v^\top=(a,b,c)$, and thus
\begin{equation} \label{eq:eqm2}
    v^\top m_{ABC}=|O_{ABC}|.
\end{equation}
Then, we can rewrite \eqref{eq:eqm1} and \eqref{eq:eqm2} as
$$\begin{pmatrix}\|p_A\|^2 \\ \|p_B\|^2 \\ \|p_C\|^2 \\ 2|O_{ABC}|\end{pmatrix}=2\begin{pmatrix}O_{ABC} & \mathbf{1} \\
    v^\top & 0\end{pmatrix}\begin{pmatrix}m_{ABC} \\ \gamma\end{pmatrix},$$
which proves \eqref{eq:lemma11}. 

Finally, to demonstrate \eqref{eq:lemma12}, we recall that for $p_i$, $i\in\{A,B,C\}$, it holds that
$$\|p_i\|^2-2p_i^\top m_{ABC}+\|m_{ABC}\|^2-r_{ABC}^2=0.$$
Also, according to \eqref{eq:eqm1}
$$\|p_i\|^2-2p_i^\top m_{ABC}-2\gamma=0,$$
then \eqref{eq:lemma12} can be inferred. 

\subsection*{Proof of Theorem \ref{thm:delaunay}}
The determinant of $M_{ABCD}$ can be computed following the Laplace expansion in the last row as 
\begin{align} \label{eq:MABCD}
    |M&_{ABCD}|= p_{x,D}\underbrace{\begin{vmatrix}p_{y,A} & p_{z,A} & 1 & \|p_A\|^2 \\ p_{y,B} & p_{z,B} & 1 & \|p_B\|^2 \\
                               p_{y,C} & p_{z,C} & 1 & \|p_C\|^2 \\ b & c & 0 & 2|O_{ABC}|\end{vmatrix}}_{|M_1|}-
                 p_{y,D}\underbrace{\begin{vmatrix}p_{x,A} & p_{z,A} & 1 & \|p_A\|^2 \\ p_{x,B} & p_{z,B} & 1 & \|p_B\|^2 \\
                               p_{x,C} & p_{z,C} & 1 & \|p_C\|^2 \\ a & c & 0 & 2|O_{ABC}|\end{vmatrix}}_{|M_2|}+ \nonumber \\
               & p_{z,D}\underbrace{\begin{vmatrix}p_{x,A} & p_{y,A} & 1 & \|p_A\|^2 \\ p_{x,B} & p_{y,B} & 1 & \|p_B\|^2 \\
                               p_{x,C} & p_{y,C} & 1 & \|p_C\|^2 \\ a & b & 0 & 2|O_{ABC}|\end{vmatrix}}_{|M_3|}-
                \underbrace{\begin{vmatrix}p_{x,A} & p_{y,A} & p_{z,A} & \|p_A\|^2 \\ p_{x,B} & p_{y,B} & p_{z,B} & \|p_B\|^2 \\
                               p_{x,C} & p_{y,C} & p_{z,C} & \|p_C\|^2 \\ a & b & c & 2|O_{ABC}|\end{vmatrix}}_{|M_4|}+\|p_D\|^2|\Lambda|.
\end{align}
The four determinants $|M_i|$, $i=1,\dots,4$ in \eqref{eq:MABCD} can be expanded again using Laplace formula in the last column. For instance, for $|M_1|$
\begin{align*}
|M_1|=&-\big(\|p_A\|^2(Adj(\Lambda))_{11}+\|p_B\|^2(Adj(\Lambda))_{12} \\
 &+\|p_C\|^2(Adj(\Lambda))_{13}+
2|O_{ABC}|(Adj(\Lambda))_{14}\big),
\end{align*}
where $(Adj(\Lambda))_{ij}$ refers to the element $(i,j)$ of the adjugate matrix of $\Lambda$. Note that the inverse matrix is defined as $\Lambda^{-1}=|\Lambda|^{-1}Adj(\Lambda)$. Similar expressions can be obtained for $|M_i|$, $i=2,3,4$, such that
\begin{align*}
    |M_{ABCD}|=&|\Lambda|\|p_D\|^2-x_D\sum_{j=1}^4(Adj(\Lambda))_{1j}(v_p)_j-y_D\sum_{j=1}^4(Adj(\Lambda))_{2j}(v_p)_j \\
    &-z_D\sum_{j=1}^4(Adj(\Lambda))_{3j}(v_p)_j -\sum_{j=1}^4(Adj(\Lambda))_{4j}(v_p)_j,
\end{align*}
where $v_p$ is defined in \eqref{eq:vp}. Thus, from \eqref{eq:lemma11}, it follows that
\begin{equation}
    |M_{ABCD}|=|\Lambda|(\|p_D\|^2-2p_D^\top m_{ABC}-2\gamma).
\end{equation}
The gradient of the determinant of $M_{ABCD}$ is
$$\nabla |M_{ABCD}|=2|\Lambda|(p_D-m_{ABC}),$$
and the Hessian matrix is
$$H(|M_{ABCD}|)=2|\Lambda|\cdot I.$$
Therefore, since $|\Lambda|$ is always negative, $|M_{ABCD}|$ is a concave function, whose maximum is at $m_{ABC}$ and $|M_{ABCD}|=0$ if $\|p_D-m_{ABC}\|=r_{ABC}$. Moreover, it holds that
$$|M_{ABCD}|\begin{cases} 
   <0 & \text{if } \|p_D-m_{ABC}\|>r_{ABC} \\
   >0       & \text{if } \|p_D-m_{ABC}\|<r_{ABC},
  \end{cases}$$
which completes the proof.

\subsection*{Proof of Lemma \ref{lem2}}
The rank of $J_{R\mathcal{S}}$ is $rank(J_{R\mathcal{S}}(p,z)) \leq   rank(R(z))+ rank(J_{\mathcal{S}}(p))$, and the kernel $ker(J_{R\mathcal{S}}) = ker(R(z))\cap ker(J_{\mathcal{S}})$ \citep{Horn2012}.
The rank of $R(z)=N_e$ since the number of edges is in the interval $[2N-2,3N-6]$ according to Proposition \ref{prop:prop1} and the graph $\mathcal{G}$ is a Delaunay triangulation to be embedded in $\mathcal{S}$ but with $z_i\in\mathbb{R}^3$. Moreover, the rank of $J_{\mathcal{S}}(p)$ is $N$ due to its block diagonal structure. Then, when the number of edges is minimal ($N_e=2N-2$), then $rank(J_{R\mathcal{S}}) \leq 2N-2+N=3N-2$. 

Moreover, we know that rigid body motions are in the kernel of $R(z)$, that is, $R(z)v=0$, such that $v^\top=(v_1^\top,\dots,v_N^\top)$ and $v_i=v_0+\omega\times p_i$, $i=1\dots,N$, where $v_0\in\mathbb{R}^3$ is a translational velocity and $\omega\in\mathbb{R}^3$ is an angular velocity. However, it is easy to see that $J_\mathcal{S}(p)(1_N\otimes v_0)\neq 0$ for $v_0\neq 0$, and that for the angular velocity it holds for each block that
\begin{equation} \label{eq:jsangular}
    p_i^\top Q_1 (\omega\times p_i)=Q_1p_i\cdot(\omega\times p_i)=\omega\cdot (Q_1p_i\times p_i),
\end{equation}
which is not zero for the general case. However, we distinguish the following cases:
\begin{itemize}
    \item If $Q_1=\alpha I_3$, $\alpha\in\mathbb{R}_{>0}$, then \eqref{eq:jsangular} is 0 $\forall \omega\in\mathbb{R}^3$, which corresponds to $s=3$ in \eqref{eq:casess}. Then, $rank(J_{R\mathcal{S}})=3N-3$ independently of the number of edges of the triangulation if $\mathcal{S}$ is symmetric in the three axes. 
    \item If $Q_1$ has some $q_i=q_j$ but $q_k\neq q_i$ for one of the axes, then \eqref{eq:jsangular} is 0 if $\omega_i=\omega_j=0$ but $\omega_k\neq 0$, That is, the components of $\omega$, corresponding to the axes in which $\mathcal{S}$ is symmetric, are zero (case $s=1$ in \eqref{eq:casess}). In that case, $rank(J_{R\mathcal{S}})=N_e+N-1$ if $N_e\leq 2N$ and $3N-1$ otherwise.
    \item If $\mathcal{S}$ is not symmetric in any of the axes, then \eqref{eq:jsangular} is not zero, and then the intersection of the kernels of $R(z)$ and $J_\mathcal{S}(p)$ is $\emptyset$ and then $rank(J_{R\mathcal{S}})=min(N_e+N,3N)$.
\end{itemize}
Then, the proof is completed.

\subsection*{Proof of Theorem \ref{thm:thm2}}
According to Lemma \ref{lemma1}, the Lyapunov function is not increasing along the systems solutions of the system and  \eqref{eq:setDesired} is an equilibrium set. Thus, the control objective is locally reached asymptotically if \eqref{eq:setDesired} is a minimum of the Lyapunov function \eqref{eq:potential}. 

Studying the Hessian matrix of a function provides information about the nature of a critical point. More specifically, if the Hessian of $W$, $H_W$, at the critical point $p^*\in\mathcal{M}_d$ is a positive-definite matrix, then $p^*$ is a local minimum. Thus, the Hessian matrix of the Lyapunov function \eqref{eq:potential} is the Jacobian of $\nabla W$. According to Lemma \ref{lemma1}, $\nabla W=\kappa_1e^\top R(z)+\kappa_2f_\mathcal{S}(p)^\top J_\mathcal{S}(p)=\xi^\top J_{R\mathcal{S}}$. Thus
$$H_W=\frac{\partial \xi^\top}{\partial p} J_{R\mathcal{S}}+\xi^\top\frac{\partial J_{R\mathcal{S}}}{\partial p}.$$
If we evaluate $H_W$ at the critical point $p^*$, i.e, when $\xi=0$, the second term is 0. Moreover, it holds that
$$\frac{\partial \xi^\top}{\partial p}=\Big(\frac{\partial e^\top}{\partial p} \ \ \frac{\partial f_\mathcal{S}^\top}{\partial p}\Big)=2\Big(R^\top(z) \ \ J_{\mathcal{S}}^\top(p)\Big)$$
Then, the Hessian at $p^*$ is
\begin{equation} \label{eq:hessianp}
    H_W(p^*)=2\begin{pmatrix} R^\top(z^*) & J_{\mathcal{S}}^\top(p^*)\end{pmatrix}\begin{pmatrix} \kappa_1R(z^*) \\ \kappa_2J_{\mathcal{S}}(p^*)\end{pmatrix}.
\end{equation}
We can define a matrix similar to $J_{R\mathcal{S}}$ but with different weights in its blocks as
$$\tilde{J}_{R\mathcal{S}}^\top(p,z)=\begin{pmatrix} \sqrt{\kappa_1}R^\top(z) & \sqrt{\kappa_2}J_{\mathcal{S}}^\top(p)\end{pmatrix}.$$
The rank of $\tilde{J}_{R\mathcal{S}}$ is the same than the rank of $J_{R\mathcal{S}}$, which is analyzed in Lemma \ref{lem2}. Thus, the Hessian matrix \eqref{eq:hessianp} can be rewritten as
\begin{equation}
H_W(p^*)=2\tilde{J}_{R\mathcal{S}}^\top(p^*,z^*)\tilde{J}_{R\mathcal{S}}(p^*,z^*).
\end{equation}
Note that $H_W(p^*)\in\mathbb{R}^{3N\times 3N}$ is positive or semipositive definite by construction since any matrix $M$ of the form $M=B^\top B$, with $B$ real, is positive or semipositive definite, and $rank(M)=rank(B)$. More specifically, if $rank(\tilde{J}_{R\mathcal{S}})=rank(J_{R\mathcal{S}})=3N$ then $H_W(p^*)$ is positive definite. According to Lemma \ref{lem2} this is the case when $N_e\geq 2N$ and the surface \eqref{eq:shieldSurface} has no symmetries. In that case, we can conclude that $p^*$ is a locally stable critical point.

We next analyze the cases in which the surface \eqref{eq:shieldSurface} has one or more symmetries (cases $s=1, 3$ in \eqref{eq:casess}) and $N_e\geq 2N$. In these cases, the dimension of the kernel of $J_{R\mathcal{S}}$ is $s$,  $H_W(p^*)$ has $s$ 0 eigenvalues and, therefore, is semi-positive definite so that we cannot conclude in principle that $p^*$ is a local minimum. However, in such case, a similar analysis can be applied as Theorem 4 in \citep{Krick2009} taking into account the following issues:
\begin{itemize}
\item Since $1\otimes v$ is not an eigenvector of $J_{R\mathcal{S}}$, the dynamics of $p$ does not contain any component that is stationary, so a reduced version of $p$ is not required.
\item The linearized dynamics of the system \eqref{eq:dynamicsAll} at $p^*$ is
$$\delta p= -H_W(p^*)p,$$
and the dynamics of $p$ near $p^*$ is
$$\dot{p}=-H_W(p^*)p-(f(p)-H_W(p^*)p)$$
where $f(p)=J_{R\mathcal{S}}(p,z)\xi(p,z)$. An orthonormal transformation $Q$ can be applied to $H_W(p^*)$ such that $QH_W(p^*)Q^\top$ is in block diagonal form with the first block of dimension $\mathbb{R}^{s\times s}$ of zeros and a second block $B\in\mathbb{R}^{(3N-s)\times (3N-s)}$ which is Hurwitz. 
\end{itemize}
Then, the center manifold theory can be applied since the system can be expressed in normal form. Finally, similar arguments follow when the number of edges is $N_e<2N$, since also the kernel of $H_W(p^*)$ will have at most dimension 3. 

\bibliographystyle{elsarticle-num-names} 
\bibliography{cas-refs}


\end{document}